\documentclass[sigconf, screen]{acmart}

\usepackage{amsmath,amssymb,amsfonts}
\usepackage{amssymb}
\usepackage{pifont}
\usepackage{algorithmic}
\usepackage{graphicx}
\usepackage{textcomp}
\usepackage{xcolor}
\def\BibTeX{{\rm B\kern-.05em{\sc i\kern-.025em b}\kern-.08em
    T\kern-.1667em\lower.7ex\hbox{E}\kern-.125emX}}
\usepackage[]{hyperref}
\usepackage{listings}
\usepackage[leftline=false,
            rightline=false,
            linewidth=.6pt,
            innerrightmargin=0pt,
            innerleftmargin=0pt,
            innertopmargin=0pt,
            innerbottommargin=0pt
]{mdframed}
\usepackage{wrapfig}
\usepackage{comment}
\usepackage{dblfloatfix}
\usepackage{subfig}
\usepackage{hyperref}
\usepackage{url}
\usepackage{color,soul}
\usepackage{xcolor}
\usepackage{soul}

\newcommand{\cmark}{\ding{51}}%
\newcommand{\xmark}{\ding{55}}%

\mdfdefinestyle{mdstyle}{
    backgroundcolor=gray!10,
    innerleftmargin=0.1cm,
    innerrightmargin=0.1cm,
    innertopmargin=0.1cm,
    innerbottommargin=0.1cm,
    leftline=true,
    rightline=true,
}
\usepackage{tikz}

\newcommand{\ignore}[1]{}

\AtBeginDocument{%
  \providecommand\BibTeX{{%
    Bib\TeX}}}

\setcopyright{none} 
\copyrightyear{2025}
\acmYear{2025}
\acmDOI{XXXXXXX.XXXXXXX}

\acmConference[]{}{}{}

\acmISBN{978-X-XXXX-XXXX-X/XX/XX}

\setcopyright{none}
\renewcommand\footnotetextcopyrightpermission[1]{}

\begin{document}

\title[DECA: A Near-Core LLM Decompression Accelerator Grounded on a 3D Roofline Model] {DECA: A Near-Core LLM Decompression Accelerator \\ Grounded on a 3D Roofline Model}

\subtitle{\hfill \break
\Large{Gerasimos Gerogiannis$^{1,3}$, Stijn Eyerman$^{1}$, Evangelos Georganas$^{2}$, Wim Heirman$^{1}$, Josep Torrellas$^{3}$}
\\ \large{
$^1$Intel Corporation \space\space\space\space\space\space\space\space\space\space\space\space $^2$Intel Labs \space\space\space\space\space\space\space\space\space\space\space\space $^3$University of Illinois at Urbana-Champaign}
\\
\vspace{-0.3cm}
\normalsize{gg24@illinois.edu, stijn.eyerman@intel.com, evangelos.georganas@intel.com, wim.heirman@intel.com, torrella@illinois.edu}}




\begin{abstract}
To alleviate the memory bandwidth bottleneck in  Large Language Model (LLM) inference workloads, 
weight matrices are stored in memory in quantized and sparsified formats. Hence, before 
tiles of these matrices can be processed by in-core generalized matrix multiplication (GeMM) hardware engines, they 
need to be dequantized and de-sparsified. This is currently performed in software with vector operations.
Unfortunately, this approach delivers only modest performance. Moreover, it is 
hard to understand how to improve the system, as the overall GeMM performance
depends on the interaction between memory resources, vector units, and
hardware matrix engines.

To improve the performance of LLM inference in advanced platforms
equipped with in-core GeMM engines and HBM, this paper makes three main contributions.
First, it develops
an analytical performance model with a 3D visual representation
that provides insights into how 
memory resources, vector units, and
hardware matrix engines interact to deliver 
compressed 
GeMM performance.
Second, it proposes {\em DECA}, a 
new near-core ML-model decompression accelerator. DECA
offloads tile 
de-sparsification and dequantization
from the
CPU, producing ready-to-use tiles for in-core GeMM engines.
Third, it introduces a new ISA extension 
that enables out-of-order invocation of the near-core accelerator.
With this extension, accelerator and core computations can interleave and overlap with high-performance.
Our evaluation shows that, in a simulated
56-core Xeon 4 server with HBM, DECA accelerates the execution of
compressed GeMMs by up to 4x over the use of optimized Intel
software kernels. Further, DECA reduces the next-token generation time of
Llama2-70B and OPT-66B by 1.6$\times$—2.6$\times$.

\end{abstract}

\maketitle

\section{Introduction}
\label{sec:intro}

Large Language Models (LLMs) are one of the most important  Machine Learning (ML) workloads, excelling at tasks such as chatbots, translation, text summarization, and content creation~\cite{zhou2024survey,zhang2023summit,kalla2023study,yao2023empowering}. LLMs use  transformers~\cite{vaswani2017attention}   
and  mainly consist of multi-head attention  and fully connected (FC) layers. The largest   models contain trillions of parameters (weights) in the FC layers~\cite{achiam2023gpt,zhao2023survey}. During  inference, these weights have low reuse (e.g., in small batch  scenarios), stressing not only the memory capacity of modern  platforms but also their memory bandwidth~\cite{yuan2024llm}.       

GPUs are regarded as the standard platform for LLM inference because of their high compute and memory bandwidth. However, recent advances introduced by   Intel Xeon 4 servers (codenamed Sapphire Rapids (SPR))~\cite{spr}, make CPUs an additional attractive option for LLM inference. First, such processors are equipped with an in-core generalized matrix multiplication (GeMM)  engine called  TMUL~\cite{intel_optimization_manual}. The TMUL serves the same purpose as the GPU Tensor Cores~\cite{markidis2018nvidia}. It is programmed with the AMX ISA extensions~\cite{intel_optimization_manual} to perform GeMMs on matrix tiles. The result is
an order of magnitude increase in GeMM computational throughput  compared to  relying solely on vector SIMD units. Second,  SPR servers can be equipped with High Bandwidth Memory (HBM), increasing the available memory bandwidth by 3-4$\times$ over  their DDR-based counterparts.

In SPR CPUs, we observe that, similar to GPUs~\cite{yuan2024llm}, LLM inference  is  memory-bandwidth bound. The large GeMMs in the FC layers account for more than 90\% of the next token generation time for LLama2-70B~\cite{llama2}. Such GeMMs have low arithmetic intensity and load a large number of weights from main memory. To a large extent, accelerating LLM inference on CPUs means speeding-up these large GeMMs.

Deep neural network (DNN) model compression techniques~\cite{liang2021pruning,deng2020model}, such as low-bit weight quantization~\cite{gholami2022survey} and sparsification/pruning~\cite{hoefler2021sparsity,xia2023flashllm,zhu2023surveyllmcompression} can improve GeMM performance: the amount of data that needs to be loaded from memory is reduced, leading to significant speedups in memory-bound kernels. Sadly, like systolic arrays~\cite{tpuv4} and Tensor Cores~\cite{xia2023flashllm}, the TMUL cannot handle arbitrary quantization schemes or sparse patterns. Consequently, the SPR TMUL engine expects well-formed dense input tiles (i.e., zero values must be included) either in BF16~\cite{kalamkar2019study} or INT8 format.

To   benefit from both model compression and TMUL GeMM throughput, Intel has recently introduced specialized kernels   in the libxsmm framework~\cite{heinecke2016libxsmm}. 
Libxsmm uses a sequence of vector (AVX) instructions to read  compressed tiles from memory, de-sparsify and/or dequantize them, and feed them to the TMUL AMX unit.
This cooperative processing mode involves two different computational domains (vector and matrix), each with its own instructions (AVX and AMX), and functional units (SIMD units and TMUL).

We profiled the performance of the libxsmm kernels for different quantized and sparsified workloads.
 Our analysis shows that, although they are very effective for moderately compressed GeMMs and with 
 the relatively low-bandwidth DDR memory, their performance degrades with HBM. This  degradation cannot be explained using a traditional two-dimensional (2D) roofline performance model~\cite{zhang2015optimizing} that only considers the memory bandwidth and the (matrix) compute throughput as bounding factors.

To guide performance optimization, we first construct an analytical performance model that 
captures the interactions between memory, matrix and vector resources. In contrast to the 2D roofline, this model has a 3D visual representation with a surface separating achievable from non-achievable performance.
For this reason, we call the model \emph{Roof-Surface}. The Roof-Surface offers useful performance insights and accurately attributes the libxsmm performance degradation to the AVX vector decompression sequence. Further, it reveals that  
overcoming the decompression inefficiencies would require a  prohibitive scaling of the CPU core's resources.

To address this problem, this paper proposes \emph{DECA}, a new
near-core {\em accelerator of ML model decompression}. DECA offloads tile
de-sparsification and   dequantization from
the CPU, producing ready-to-use tiles for the TMUL. 
DECA can be programmed to 
 handle quantized number formats with any number of bits between 1 and 8,
supports any level of unstructured sparsity, and supports group quantization~\cite{gholami2022survey}.
The DECA microarchitecture
performs decompression by utilizing a   pipeline with advanced vector operations. Importantly, we use the {\em Roof-Surface} model, (1) to make decisions about the vector pipeline  microarchitecture and (2) to perform design space exploration and derive a well-balanced DECA design.

We observe that if the CPU cores use regular memory-mapped load/store instructions to communicate with DECA, the communication latency gets exposed and hurts performance. 
To this end, 
we introduce a new ISA extension that hides the CPU-DECA communication latency by invoking the accelerator out-of-order.
We call this  extension 
{\em Tile External Preprocess and Load} (TEPL).

Our  evaluation for two different low-bit quantization formats (BF8 and MXFP4) and different unstructured sparsity levels shows that DECA is very effective. In a 
simulated 56-core SPR with HBM, DECA accelerates the execution of
compressed GeMMs by up to 4x over the optimized Intel libxsmm software kernels.
In addition, by speeding-up the  FC layers, DECA reduces the next-token generation time of Llama2-70B and OPT-66B~\cite{zhang2022opt} by 1.6$\times$—2.6$\times$ over  the software-only solution, and by 2.5$\times$—5.0$\times$ over the uncompressed baseline model.

This paper's contributions are:

\noindent $\bullet$ The \emph{Roof-Surface} performance model that models the interaction between vector units, matrix units, and memory.

\noindent $\bullet$ The \emph{DECA} near-core accelerator, designed to 
 accelerate the de-sparsification and dequantization of compressed ML models.
 
\noindent $\bullet$ The Tile External Preprocess \& Load (\emph{TEPL}) extension that enables out-of-order invocation of near-core accelerators.

\noindent $\bullet$ A simulation-based evaluation of the performance of \emph{DECA} for compressed GeMMs in LLM inference.

\vspace{-0.2cm}

\section{Background}
\label{sec:background}
\vspace{-0.025cm}
\subsection{LLM Inference}
\vspace{-0.05cm}

Large Language Models (LLMs) consist of different layers, such as Embedding layers, Fully-Connected (FC) layers and Attention layers~\cite{vaswani2017attention}. 
LLM inference has two phases~\cite{patel2023splitwise}. 
The first one encodes the input tokens and generates the first token (prompt phase). The second one generates the next output tokens (generation phase).
In this work, we focus on executing the low arithmetic-intensity generation phase efficiently, since for many practical use cases it dominates the end-to-end LLM inference time~\cite{yuan2024llm}.

GPUs are regarded as the standard platform for LLM inference~\cite{patel2023splitwise,su2023synergy} because of their high compute and memory bandwidth. However, recent advances, such as HBM and in-core GeMM engines~\cite{spr}, make CPUs
an additional attractive option for LLM inference. There has been increasing research and industrial interest in making CPUs better at Machine Learning (ML) and scientific workloads, by either incorporating extensions or small accelerators on the CPU die ~\cite{jeong2023vegeta,save,graphite,tmu,nassif2022sapphire,orenes2022tiny}. For these reasons, in this paper, we focus on LLM inference on modern CPU servers.

\subsection{Model Compression} \label{subsec:model_compr}

For low arithmetic-intensity LLM FC layers, compressing the weight matrices reduces data movement  and, therefore, can directly improve performance in both GPUs and CPUs.
There are two main ways to compress an ML model~\cite{han2015deep,zhu2023surveyllmcompression}:

\noindent $\bullet$ \textbf{Quantization} involves storing weights in a lower-bit format, e.g., FP8 or FP4 instead of FP16. Multiple quantization schemes exist~\cite{awq,zhao2024atom,kim2024memory,wei2023outlier}. Some of them additionally split weights in groups and introduce a per-group scaling factor (\emph{group quantization}) to achieve higher accuracy. We evaluate two types of weight quantization in our work: BF8 (8-bit brain floating point) and MXFP4~\cite{mxfp4}. The latter uses a 4-bit floating point
and group quantization with a shared scaling factor for every 32 weights (8 bit exponent); it has been shown to not degrade LLM accuracy~\cite{mxfp4}.

\noindent $\bullet$ \textbf{Sparsification} consists of eliminating (\emph{pruning}) weights that are close to zero and/or that do not contribute much to the model's accuracy~\cite{lecun1989optimal,blalock2020state,hoefler2021sparsity}. \emph{Unstructured sparsity} does not impose restrictions on which weights can be removed. It achieves higher accuracy 
than structured sparsity for the same sparsity level~\cite{liu2021group,sparsegpt}. In this work, we assume unstructured sparsity and a bitmask-based sparse format, to avoid storing zeros. To reconstruct the position of the non-zeros in the original weight matrix, a bitmask is used that has as many bits as the number of elements in the original matrix.
The '1' bits in the bitmask indicate the location of the nonzeros, which are stored consecutively in 
a nonzero array. Recently proposed LLM weight pruning methods  such as SparseGPT~\cite{sparsegpt} have achieved unstructured sparsity levels of up to 60-70\% without significant loss in accuracy. For traditional ML models like ResNet50~\cite{resnet50}, unstructured sparsity levels up to 95\% are easy to achieve~\cite{peste2021ac}. Since we believe that LLM research advances will soon enable higher sparsity levels, we evaluate a large 50\%--95\% range of sparsity.

Models may be both sparse and quantized~\cite{harma2024effective}. 
Starting from a dense BF16 model, a $Q$ bit quantized model with a density factor of $d$ (e.g., $d=10\%$ means only 10\% of the weights are nonzeros), reduces the model size by a factor of $16/(Q\times d+1)$, where the '1' comes from the bitmask bit. We assume that the footprint of activations is negligible. We refer to this factor as \emph{Compression Factor (CF)}.

The compression process is executed offline (e.g., after training). It is shown on the left part of Figure~\ref{fig:compress_decompress}.
In this paper, we assume an already compressed model that we want to use online for inference.

\vspace{-0.3cm}

\subsection{Matrix Extensions} \label{subsec:amx}

There are several matrix extensions~\cite{amx,arm_sme,de2022compiling,bhat2021matrix}
to improve the efficiency of matrix multiplication on CPUs. In this work, we use Intel's Advanced Matrix Extensions (AMX)\cite{amx}.
AMX extends the register file with 8 matrix registers, called tile registers.
Each one can hold up to 16 rows, with 64 bytes of data per row that can be interpreted as 32 2-byte elements (BF16) or 64 1-byte elements (INT8).
Each tile 
can contain up to 1 KB of data.

Each core has the tile registers and a matrix multiplication TMUL unit that multiplies the tiles.
To load/store data to/from the tile registers, AMX includes tload/tstore instructions.
For the next token generation phase in LLMs with a batch size of $N\leq$16  and BF16,
a weight tile $W$ contains $M=16$ rows each with $K=32$ columns. An activation tile $A$ contains $N$ rows each with $K=32$ columns. The TMUL performs the operation $A\times W^T$ to produce an $NxM$ output tile. The TMUL operation takes 16 cycles to execute regardless of the N value. and performs a total of $N\times K\times M =N\times32\times16=512N$ fused multiply-adds (FMA) -- or equivalently $32N$ FMAs per cycle. For N>16, the TMUL throughput saturates at 512 FMAs per cycle, since the activation tile can hold no more than 16 rows.
Any mention of FLOPs in this work refers to FMAs.

\subsection{GeMM Decompression} \label{subsec:libxsmm}

\begin{figure}[htb]
\centering
\vspace{-0.2cm}
\begin{minipage}[T]{0.55\linewidth}
    \includegraphics[width=0.9\linewidth]{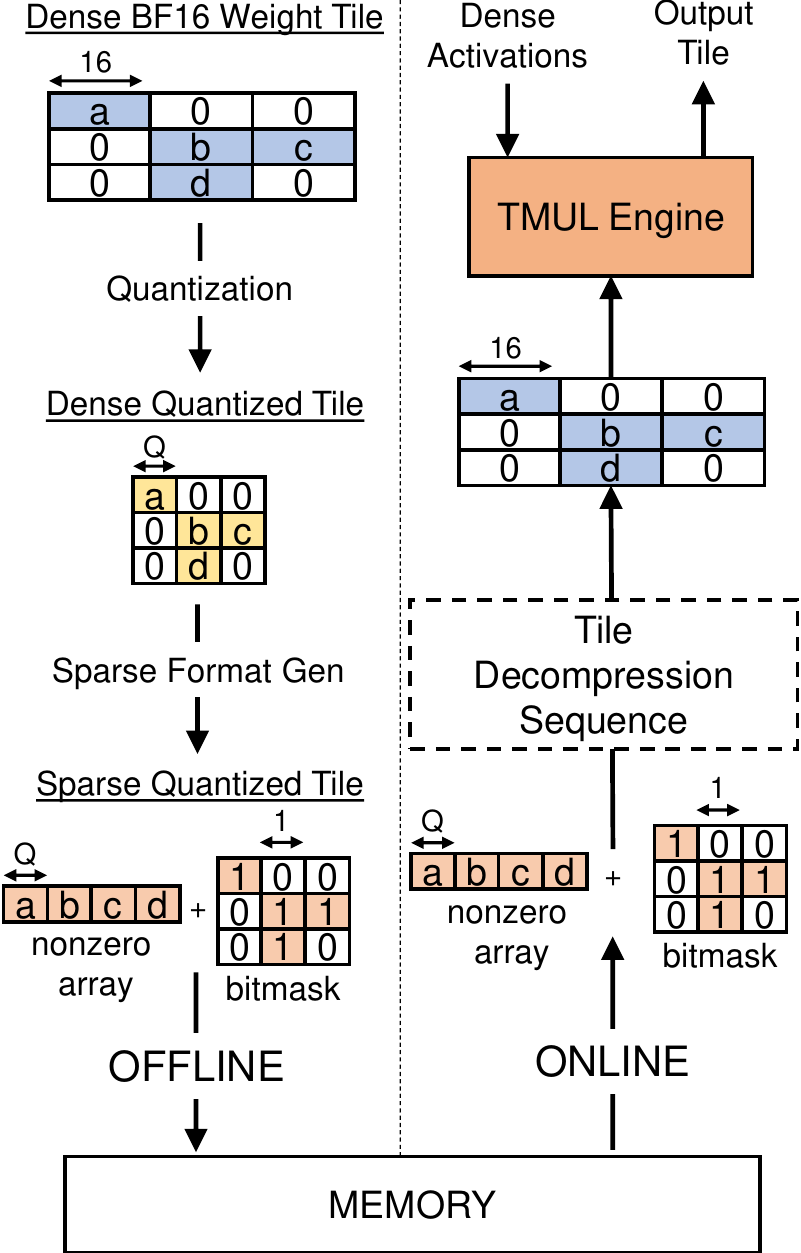}
    \vspace*{-0.35cm}
    \caption[]{Weight offline compression and online decompression.} 
    \label{fig:compress_decompress}
    \vspace{-5mm}
\end{minipage} \hspace{1mm}
\begin{minipage}[T]{0.41\linewidth}
    \begin{mdframed}
        \begin{lstlisting}
............
//Decompress Ti+1
for(r=0 to 16):
{
//Decompress
//row r of Ti+1
VectorOps AVX
...
}

//GeMM Ti
MatrixOps AMX
TLoad Ti
TComp Tout, Ti

//Decompress Ti+2
//GeMM Ti+1
............
        \end{lstlisting}
    \end{mdframed}
    \vspace*{-0.3cm}
    \caption{Libxsmm compressed GeMM kernel pseudocode.}
    \label{lst:libxsmm}
    \vspace{-0.2cm}
\end{minipage}

\end{figure}

The TMUL, similar to other GeMM engines~\cite{markidis2018nvidia,tpuv4}, can handle data in very specific data formats (i.e., BF16 or I8) and cannot handle unstructured sparsity. If a 
GeMM contains compressed weights, decompression is needed  to produce tiles that conform to the TMUL requirements. Unlike compression, decompression is performed online (Figure~\ref{fig:compress_decompress}). Thus, it can impact performance.

To achieve high performance in compressed GeMM kernels and hide the decompression overhead, Intel recently introduced a software solution integrated in the Libxsmm framework~\cite{heinecke2016libxsmm} (Figure~\ref{lst:libxsmm}). The decompression sequence is handled using  AVX vector operations, while the actual GeMM is executed using  AMX matrix operations. Libxsmm adopts a smart method to overlap the execution of the two: software allocates a double software buffer, and tries to keep it in the L1 cache. The output of the AVX decompression sequence for tile \emph{Ti+1} is written in one of the two software buffers. At the same time, AMX instructions load data from the other software buffer that contains \emph{Ti}, which has been previously decompressed by the AVX sequence. Overlapping AMX with AVX is enabled by out-of-order execution and dependencies are naturally honored.

The decompression sequence uses vector operations such as 
permutes for the decompression, and masked vector expands to insert zeros in the appropriate positions of the nonzero array. Although we omit the specifics due to limited space, 
the first takeaway is that decompression is done using AVX, utilizing a different "domain" (i.e., separate instructions and functional units) than AMX. The second takeaway is that the AVX dynamic instructions vastly outnumber the AMX ones, since AMX uses tile-sized operands (1KB), while AVX operates on cache-line sized ones (64B, one tile row).

\section{Motivation}
\label{sec:motivation}

\vspace{-0.05cm}
\subsection{GeMMs in FC Layers Dominate  Inference}

Table~\ref{tab:percentages} shows the fraction of the next-token (i.e., generation)
time spent in the GeMMs of the different Fully Connected (FC) layers of 
Llama2-70B~\cite{llama2}
on an SPR server with either DDR5 or   HBM. We show results for an uncompressed model with BF16 weights, with different numbers of input tokens and batch sizes ($N$). 
The rest of the time is spent on kernels such as attention,
for which weight compression does not apply.
We see that the  time  spent in such GeMMs is over 95\% for DDR5 and 85--90\% for HBM. 
Hence,  accelerating these GeMMs can greatly improve the next-token time.

\begin{table}[htb]
\vspace{-0.2cm}
\centering
\footnotesize
\caption{Contribution of the GeMMs of FC layers to the next-token time.}
\vspace{-0.4cm}
\label{tab:percentages}
\begin{tabular}{|r|rr|rr|}
\hline
\multicolumn{1}{|c|}{Memory} & \multicolumn{2}{c|}{DDR (260GB/s)} & \multicolumn{2}{c|}{HBM (850GB/s)}\\ \hline
\multicolumn{1}{|c|}{Input Tokens} & \multicolumn{1}{c}{32} & \multicolumn{1}{c|}{128} & \multicolumn{1}{c}{32} & \multicolumn{1}{c|}{128}\\ \hline
Batch size (N): 1 & 97.4\% & 97.5\% & 89.8\% & 89.5\% \\
4 & 97.3\% & 97.1\% & 89.4\% & 88.9\% \\
16 & 96.6\% & 95.5\% & 88.3\% & 85.9\% \\ \hline
\end{tabular} 
\vspace{-0.4cm}
\end{table}

\vspace{-0.05cm}
\subsection{GeMMs in FC Layers are Bandwidth Bound}

\begin{figure}[htb]
 \captionsetup[subfloat]{captionskip=0.7pt}
    \vspace{-0.6cm}
    \centering
  \subfloat[56-core SPR with DDR5]{%
      \includegraphics[width=0.485\linewidth]{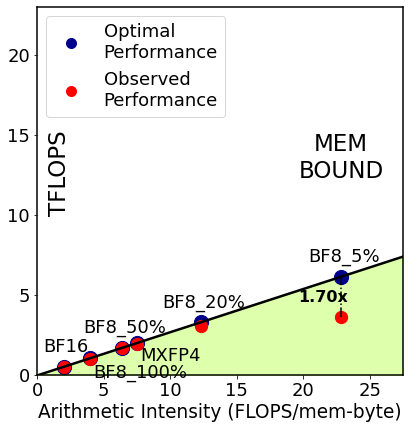}
      }
  \subfloat[56-core SPR with HBM]{%
      \includegraphics[width=0.48\linewidth]{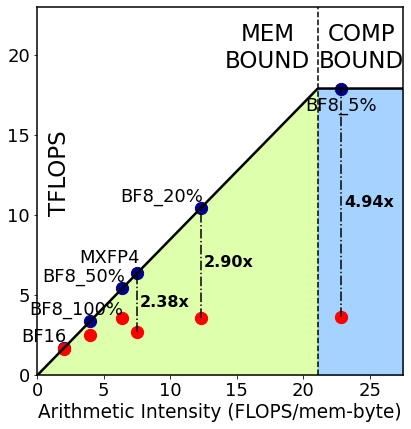}
      }
      \vspace*{-0.35cm}
  \caption{Traditional  rooflines for a GeMM with N=4.}
  \label{fig:naive_rooflines}
  \vspace{-0.5cm}
\end{figure}

Figure~\ref{fig:naive_rooflines} shows the roofline models for one of the large GeMMs of the FC layers in LLama2-70B for an SPR with either DDR5 or HBM, and  N=4.
We use the TMUL FLOPS limit (Section~\ref{subsec:amx}) for the maximum achievable GeMM FLOPS in the compute-bound area.
In this work, when calculating the Arithmetic Intensity (AI) in FLOPs per memory byte, we assume that the footprint of the weight matrices is much larger than that of activations, which is true for small values of N. 
The leftmost circle in both graphs, labeled as `BF16', is our baseline uncompressed execution.
We see that this execution is memory-bandwidth bound in both cases due to a low AI. This motivates model compression,  to reduce the amount of data that needs to be read from memory.

\subsection{Compressed GeMMs Can Introduce Inefficiencies}

The other data points in Figure~\ref{fig:naive_rooflines} represent compressed models with 
4-bit quantization (MXFP4) or with 8-bit quantization (BF8) with density levels (i.e., fraction of
nonzeros)   ranging from 5\% to 100\%.
Compression reduces the amount of data   fetched from memory, which increases the AI, moving the circles to the right as the compression factor increases. For each design point, we show two circles: one
at the {\em Observed} performance, and one on the roofline for the same AI. We call the latter {\em Optimal} performance.

We see that, as we increase the compression factor, the Observed and Optimal points increasingly
diverge. In the DDR5 graph, the divergence appears at BF8 with 5\% density. However, in the 
HBM graph, all the compressed models are below their Optimal performance; at BF8 with 5\% density,
the ratio between Optimal and Observed performance is 4.94x. This means that performance is limited 
by some inefficiency that is not captured by the roofline model. 
By manual profiling, we find that the root cause is the overhead of 
the AVX decompress instruction sequence. Effectively,
the AVX SIMD processing units of the cores are unable to keep up with the memory bandwidth 
and/or the throughput of TMUL.

Considering the importance of LLM workloads,
some form of hardware support for
the decompression overhead could be justified.
However, one has to be cautious when making changes in the resource-constrained CPU setting. The 
roofline model 
does not 
inform us on the required vector throughput improvement 
for the kernels to shift from being bounded by vector processing to  being  bounded by memory or matrix computation. There is a danger of constructing hardware solutions that are either underprovisioned or overprovisioned. 
To avoid this danger, in the next section, we propose an alternative analytical model. This model can theoretically guide the required hardware support for eliminating the decompression overhead in compressed GeMMs.

\section{The  Roof-Surface Model} 
\label{sec:roofsurface}

To guide performance optimization for our kernels that involve matrix, vector, and memory operations,  we 
develop a performance model  that captures their interaction. 
This model, which we call \emph{Roof-Surface}, has a
three-dimensional (3D) visualization.  We also
present a 2D projection, called the \emph{Bounding Region Diagram} (BORD).

\vspace{-0.15cm}

\subsection{The 3D Roof-Surface Performance Model}

In cases where multiple interacting factors can affect performance, the
slowest factor ends-up determining the performance. Thus, we should first express how fast 
(1) memory  can provide compressed tiles (MEM), 
(2) vector hardware  can process compressed tiles (VEC), and 
(3) matrix hardware can process decompressed tiles (MTX).

\vspace{1mm}
\noindent {\bf Memory}  can provide compressed tiles at a rate of $MBW/Bytes_{tile}$ tiles per second, where $MBW$ is the memory bandwidth and $Bytes_{tile}$ is the number of bytes in 
a  compressed tile. 
Since a compressed weight tile will be used for a single TMUL matrix operation, we refer to $1/Bytes_{tile}$ as matriX-to-Memory arithmetic intensity or $AI_{XM}$. It expresses how many matrix operations can be executed per byte loaded from memory, and it is very similar to the traditional arithmetic intensity used in the rooflines of Figure~\ref{fig:naive_rooflines}. The main difference is that its units are matrix operations per byte and not FLOPs per byte. In our setting, compression schemes with higher compression factors (CF) (Section~\ref{subsec:model_compr}) have a higher $AI_{XM}$. Overall, the MEM rate in compressed tiles per second is $MBW * AI_{XM}$.

\vspace{1mm}
\noindent {\bf The Vector Hardware}   decompresses tiles at a rate of $VOS/VO_{tile}$, where \emph{VOS} is the number of vector operations per second that can be executed by the architecture, and $VO_{tile}$ is the number of vector operations needed per tile. 
\emph{VOS} is the vector throughput and is an architecture-dependent parameter. For example, for our SPR system,  it is given by the product of the processor frequency (\emph{f}), the number of cores ($c$), and the number of SIMD units per core. 
$VO_{tile}$ is a kernel-dependent parameter. Since only the weight matrix in a GeMM needs to
be decompressed, $VO_{tile}$
effectively expresses how many vector operations are needed per matrix operation. We refer to 
$1/VO_{tile}$ as the matriX-to-Vector arithmetic intensity or 
$AI_{XV}$, since  it expresses how many matrix operations can be executed per vector operation. 
Overall, the VEC rate in  tiles per  second is  $VOS*AI_{XV}$.

\vspace{1mm}
\noindent {\bf The Matrix Hardware} can perform \emph{MOS} matrix operations per second. 
\emph{MOS} depends   on the architecture and not on the   kernel. For example, in SPR systems, it is given by $f*c/16$, since each core has a TMUL that
 takes 16 cycles to perform a tile multiplication. 
Overall, the MTX rate in tiles per second is simply \emph{MOS}.

\vspace{1mm}
\noindent {\bf The Final Performance} is determined by the lowest tile processing rate
among the three rates considered. Specifically, the number of tiles per second (\emph{TPS})
that the architecture can process is:
\vspace{-0.1cm}
\begin{equation}
\label{equation:tps}
    TPS=min\{MBW*AI_{XM},VOS*AI_{XV},MOS\}
\end{equation}

We can easily get the rate of FLOPs per second ({\em FLOPS}) by recalling from Section~\ref{subsec:amx} that a TMUL tile  operation corresponds to 512$*N$ FMAs. Thus:

\vspace{-0.3cm}
\begin{equation}
\label{equation:roofsurface}
FLOPS=512*N*min\{MBW*AI_{XM},VOS*AI_{XV},MOS\} 
\end{equation}

We call this equation the \emph{Roof-Surface} equation. 
Any of the three terms inside the \emph{min} clause can be the one limiting performance.
For a given architecture (i.e., fixed \emph{MBW},  \emph{VOS}, and \emph{MOS}), there are \emph{two kernel-dependent variables} 
inside the \emph{min} clause:  $AI_{XM}$ and $AI_{XV}$. 
These are the kernel's ``signature''---if two kernels have the same signature, they  have the same projected performance.
In contrast, in the roofline model, the kernel signature is just one variable: the traditional FLOP-to-memory AI. Now, the illustration of the performance model 
can no longer be done in the two dimensions of Figure~\ref{fig:naive_rooflines} (FLOP-to-memory AI and FLOPS). We need three dimensions: one for $AI_{XM}$ (x dimension), one for $AI_{XV}$ (y dimension), and one for   FLOPS (z dimension).  

\begin{figure}[htb]
    \vspace{-0.7cm}
    \centering
  \subfloat[]{%
      \includegraphics[width=0.39\textwidth]{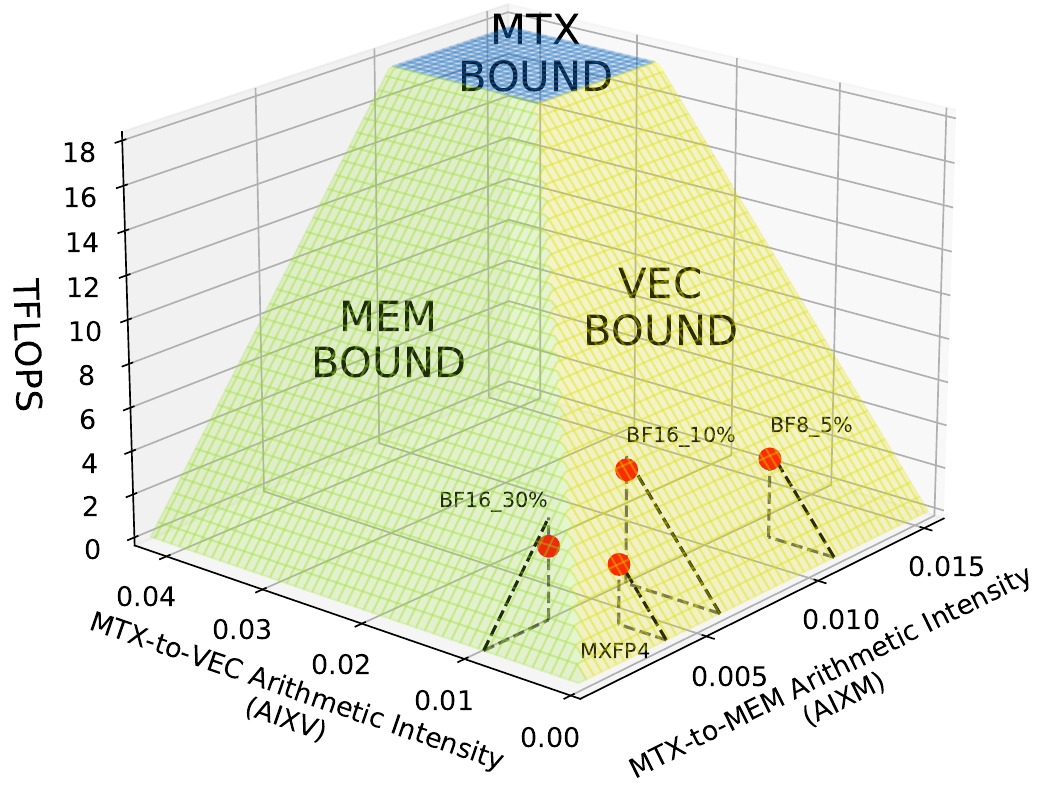}
      \label{fig:roofsurface_triangles}
      }
  \subfloat[]{%
\resizebox{0.09\textwidth}{!}{

\begin{tabular}[b]{lll}
\hline
\multicolumn{1}{|l|}{\textbf{R-L}}  & \multicolumn{1}{l|}{\textbf{R-S}} & \multicolumn{1}{l|}{\textbf{Real}} \\ \hline
\multicolumn{3}{c}{MXFP4}                                                         \\ \hline
\multicolumn{1}{|l|}{6.3}  & \multicolumn{1}{l|}{2.9} & \multicolumn{1}{l|}{2.7}  \\ \hline
\multicolumn{3}{c}{BF8}                                                           \\ \hline
\multicolumn{1}{|l|}{3.3}  & \multicolumn{1}{l|}{3.3} & \multicolumn{1}{l|}{2.5}  \\ \hline
\multicolumn{3}{c}{BF8\_50\%}                                                     \\ \hline
\multicolumn{1}{|l|}{5.3}  & \multicolumn{1}{l|}{4.0} & \multicolumn{1}{l|}{3.6}  \\ \hline
\multicolumn{3}{c}{BF8\_30\%}                                                     \\ \hline
\multicolumn{1}{|l|}{7.8}  & \multicolumn{1}{l|}{4.0} & \multicolumn{1}{l|}{3.6}  \\ \hline
\multicolumn{3}{c}{BF8\_20\%}                                                     \\ \hline
\multicolumn{1}{|l|}{10.2} & \multicolumn{1}{l|}{4.0} & \multicolumn{1}{l|}{3.6}  \\ \hline
\multicolumn{3}{c}{BF8\_10\%}                                                     \\ \hline
\multicolumn{1}{|l|}{14.8} & \multicolumn{1}{l|}{4.0} & \multicolumn{1}{l|}{3.6}  \\ \hline
\multicolumn{3}{c}{BF8\_5\%}                                                      \\ \hline
\multicolumn{1}{|l|}{17.5} & \multicolumn{1}{l|}{4.0} & \multicolumn{1}{l|}{3.6}  \\ \hline
\multicolumn{3}{c}{BF16\_50\%}                                                    \\ \hline
\multicolumn{1}{|l|}{3.0}  & \multicolumn{1}{l|}{3.0} & \multicolumn{1}{l|}{2.5}  \\ \hline
\multicolumn{3}{c}{BF16\_30\%}                                                    \\ \hline
\multicolumn{1}{|l|}{4.6}  & \multicolumn{1}{l|}{4.6} & \multicolumn{1}{l|}{3.3}  \\ \hline
\multicolumn{3}{c}{BF16\_20\%}                                                    \\ \hline
\multicolumn{1}{|l|}{6.3}  & \multicolumn{1}{l|}{5.7} & \multicolumn{1}{l|}{4.2}  \\ \hline
\multicolumn{3}{c}{BF16\_10\%}                                                    \\ \hline
\multicolumn{1}{|l|}{10.2} & \multicolumn{1}{l|}{5.8} & \multicolumn{1}{l|}{5.2}  \\ \hline
\multicolumn{3}{c}{BF16\_5\%}                                                     \\ \hline
\multicolumn{1}{|l|}{14.8} & \multicolumn{1}{l|}{5.8} & \multicolumn{1}{l|}{5.5}  \\ \hline
\end{tabular}

    \label{fig:roofsurface_accuracy}
}}

    \vspace*{-0.3cm}
  \caption{(a) The 3D Roof-Surface model. (b) The optimal performance based on
  the Roofline (R-L), the Roof-Surface (R-S), and  real performance measurements in TFLOPs.}
  \label{fig:roofsurface_viz_imprv}
  \vspace{-0.35cm}
\end{figure}

Figure~\ref{fig:roofsurface_triangles} shows the result of plotting Equation~\ref{equation:roofsurface} (for N=4,HBM) in three dimensions to form the {\em Roof-Surface} plot. A  Roof-Surface  plot 
has three regions, depicted in different colors. In each of the regions, a different term of 
the Roof-Surface Equation is the smallest one, and thus bounds performance.
The operation points below the blue subsurface are bound by the MTX factor, 
the ones below the green subsurface are bound by the MEM factor, and the ones 
below the orange subsurface are bound by the VEC factor. 
Kernel performance is depicted by points in the 3D space. 
The achievable performance
is bounded by the overall surface, rather than by a line like in the roofline model.
For this reason, we call the model  Roof-Surface.
Points above
the overall surface are not achievable.

Figure~\ref{fig:roofsurface_triangles} also includes red points that correspond to the observed performance points for different compression schemes. 
We see that the red points under the VEC-bound region (MXFP4, BF16\_10\%, BF8\_5\%) are very near to the top of the corresponding tangent triangles (i.e. almost exactly on the roofsurface). 
This visually reveals that they are bounded by vector operations. The red point in the MEM-bound region (BF16\_30\%) is slightly below the roofsurface, revealing that, for this point, a non-plotted factor such as memory latency is leaving a little bit of performance on the table.

In Figure~\ref{fig:roofsurface_accuracy} we show the optimal performance values as predicted by the roofline (R-L) and the Roof-Surface (R-S) models, and the real
observed values. For almost all kernels, the Roof-Surface produces accurate performance bounds, while the roofline can be way off. If we were to plot many of the roofline predictions on the 3D space they would float above the roofsurface.
Note that for kernels BF8, BF16\_50\%, and BF16\_30\%, 
the performance estimates of R-L and R-S are the same. The reason is that these kernels are classified as MEM-bound by both models.

\vspace{-0.2cm}
\subsection{The 2D Bounding Region Diagram}

We introduce an easier to visualize 2D representation of the Roof-Surface plot that we
call the Bounding Region Diagram (BORD). BORD is the projection of the roofsurface on the xy plane.  A BORD does not depict FLOPS information, but accurately identifies which one of the plotted factors bounds the performance of a given kernel.

\begin{figure}[htb]
    \vspace{-0.6cm}
    \centering
  \subfloat[2D BORD for HBM SPR]{%
      \includegraphics[width=0.238\textwidth]{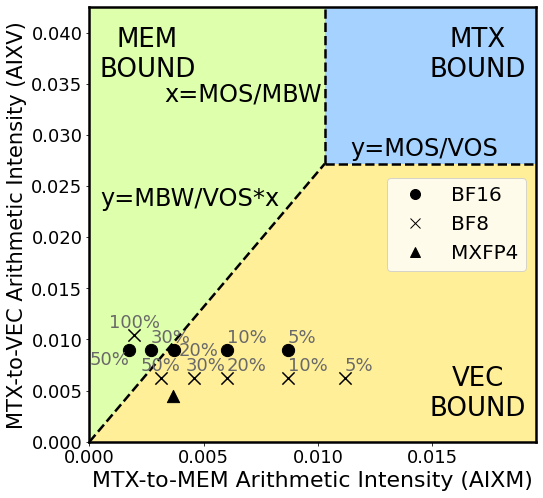}
      \label{fig:bored_hbm}
      } 
  \subfloat[2D BORD for DDR SPR]{%
      \includegraphics[width=0.238\textwidth]{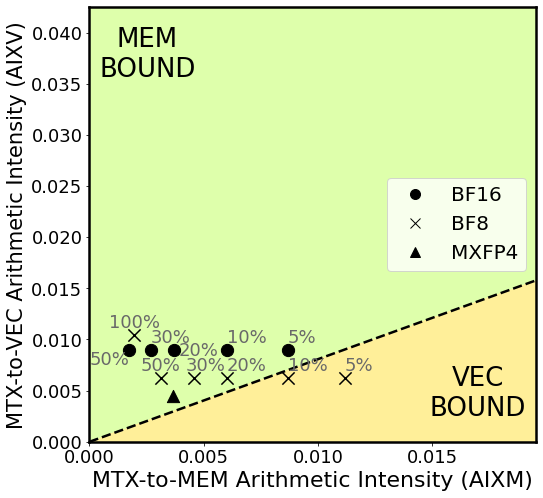}
      \label{fig:bored_ddr}
      }
      \vspace*{-0.3cm}
  \caption{2D bounding-region diagrams (BORD).}
  \label{fig:roofsurface}
  \vspace{-0.35cm}
\end{figure}

Figure~\ref{fig:bored_hbm} shows the BORD for  HBM SPR.
The figure shows the equations of the lines that separate the three
regions. They are:  $y=(MBW/VOS)*x$, $x=MOS/MBW$, and $y=MOS/VOS$. It also shows 
the positions of the different  
compressed GeMM kernels that use BF8 and MXFP4 from Figure~\ref{fig:naive_rooflines}b,
and of
additional   kernels that use  
BF16 with different density levels.
We observe that the  vast majority of kernels are VEC-bound. 
To reach the 
performance  of the roofline in Figure~\ref{fig:naive_rooflines}b, these points should be pushed away from the VEC-bound region.

Figure~\ref{fig:bored_ddr} shows the  BORD for  DDR SPR, which has a smaller \emph{MBW} value. Now, the area of the MEM-bound region increases.
The MTX-bound  region is no longer visible for the $AI_{XM}$ and $AI_{XV}$ value ranges we are plotting in the BORD. Its area is consumed by the MEM region. The BORD also shows that all of our
kernels except BF8 with 20\% and lower density are in the MEM-bound area or very close to it. This explains that
the software decompression solution reaches the roofline  in most design points of Figure~\ref{fig:naive_rooflines}a.

\newlength{\oldintextsep}
\newlength{\oldcolumnsep}
\setlength{\oldintextsep}{\intextsep}
\setlength{\oldcolumnsep}{\columnsep}
\setlength{\intextsep}{0pt}
\setlength{\columnsep}{8pt}

\begin{wrapfigure}{r}{0.24\textwidth}
    \centering
    \includegraphics[width=\linewidth]{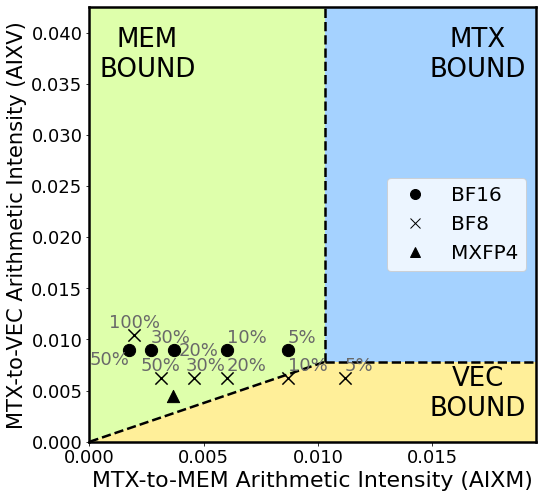}
    \vspace{-0.75cm}
    \caption{2D BORD for HBM with 4x VOS.}
    \label{fig:bored_hbm_vos}
    \vspace{-0.4cm}
\end{wrapfigure}

Finally, Figure~\ref{fig:bored_hbm_vos} shows the BORD when we take the HBM SPR variant
and increase the vector  throughput in VOS  by 4x,   in an attempt to eliminate the vector bottleneck. When compared to Figure~\ref{fig:bored_hbm}, we see that the area of the VEC-bound region decreases and the MEM-bound region covers more kernels. However, even a 4x VOS increase is not enough to make all kernels not VEC-bound.

We found that in the HBM SPR variant of Figure~\ref{fig:bored_hbm},
cores typically spend
over 95\% of their dynamic instructions on tile decompression, and that 
cores are already
using 40--80\% of their
commit slots. 
Hence, increasing the VOS by 4x would require not only a 4x increase in the number of
SIMD AVX units, but also a prohibitive increase in the core's superscalar width. We further discuss the limitations of this and other conventional solutions (such as increasing the vector width without increasing the number of AVX units) in Section~\ref{sec:alternatives} and evaluate those limitations in Section~\ref{sec:evaluation}.

\setlength{\intextsep}{\oldintextsep}
\setlength{\columnsep}{\oldcolumnsep}

\section{DECA Overview and Out-of-Order Invocation}
\label{sec:integration}

The previous analysis
reveals that, to hide the decompression overheads with a conventional solution, 
one would need a very expensive scaling of the general-purpose core's resources. 
This motivates us to propose {\em DECA}, a {\em near-core decompression accelerator for ML models}. DECA offloads vector processing for decompression from the cores. In this section,
we first describe 
DECA's integration. We then introduce a new mechanism and ISA extensions for efficiently overlapping the operation of CPU cores and near-core accelerators.

\subsection{DECA Placement \& System  Integration}

\begin{figure}[h!]
    \vspace{-0.05cm}
    \centering
    \includegraphics[width=0.9\linewidth]{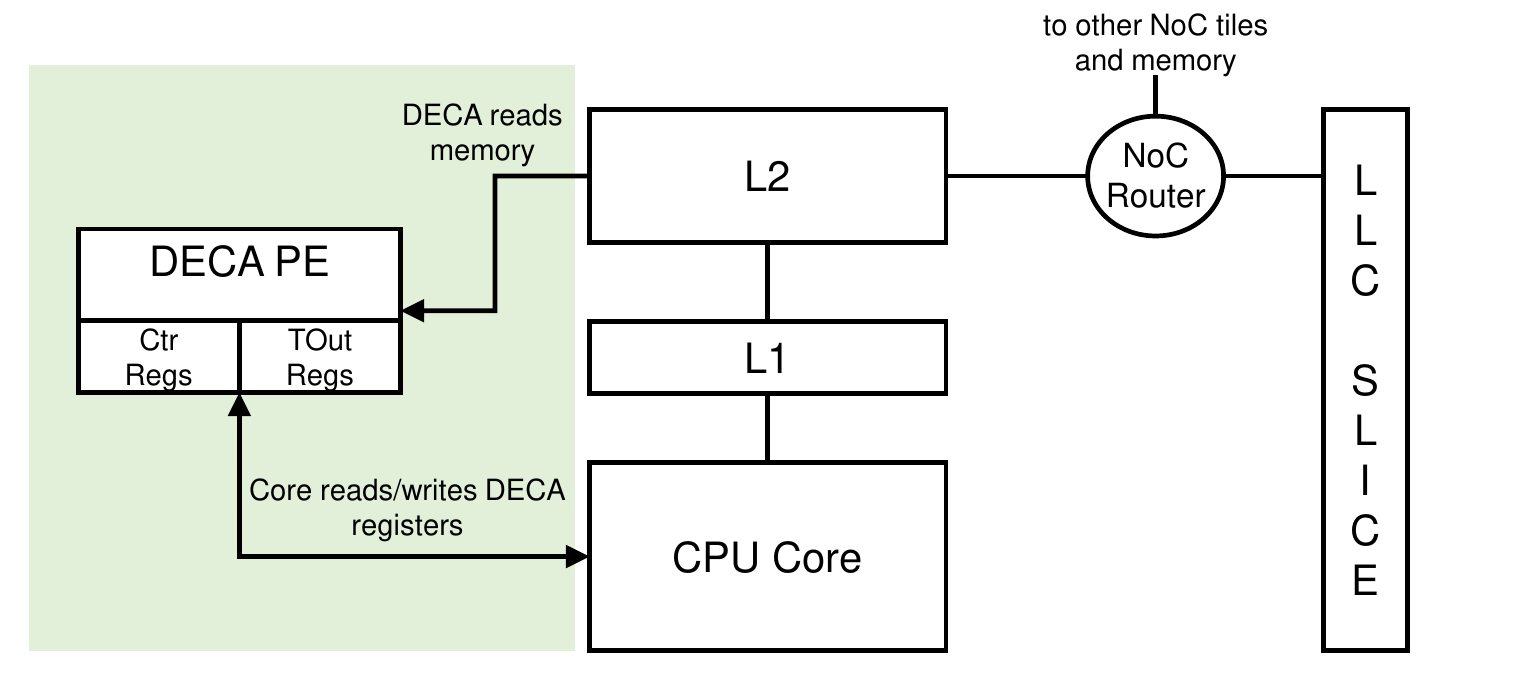}
    \vspace*{-0.45cm}
    \caption{DECA placement next to a core.}
    \label{fig:deca_placement}
   \vspace{-0.3cm}
\end{figure}

We envision a processor to have a DECA   associated with each core
as shown in Figure~\ref{fig:deca_placement}. 
A DECA has 
a memory-mapped interface that allows the core to write commands and read data.
A DECA has a processing element (PE), control registers, and tile output ({\em TOut})
registers.
The core uses privileged stores to the control registers to configure the PE to
perform decompression of tiles with a given quantization scheme and with or without sparsity. 
A configuration includes filling look-up tables (LUTs) that DECA employs for efficient dequantization (Section~\ref{sec:microarch}). 

The DECA PE reads a compressed tile from memory, processes it, and then writes the decompressed tile 
to the  TOut registers. Then, the CPU core reads the TOut registers
and uses the data to execute  
the GeMM using AMX instructions. 
The PE accesses memory through the L2, issuing both regular loads (but never stores) and prefetch requests, 
generated by a prefetcher integrated in the PE.
DECA shares the L2 TLB with the core like prior
work~\cite{graphite,gerogiannis23spade,tmu} 
and, therefore, uses the virtual space of the CPU core.

A DECA can potentially be used by multiple processes. One approach is to save
and restore the DECA state on context switches. 
Alternatively, 
we propose that DECA retains its
 state across context switches and  when a new process attempts to use DECA, it causes a trap to
the OS, which saves the state and reconfigures DECA.

\vspace{-0.1cm}
\subsection{DECA-Core Cooperative Tile Processing}

\begin{figure}[h]
    \vspace{-0.3cm}
    \centering
    \includegraphics[width=0.9\linewidth]{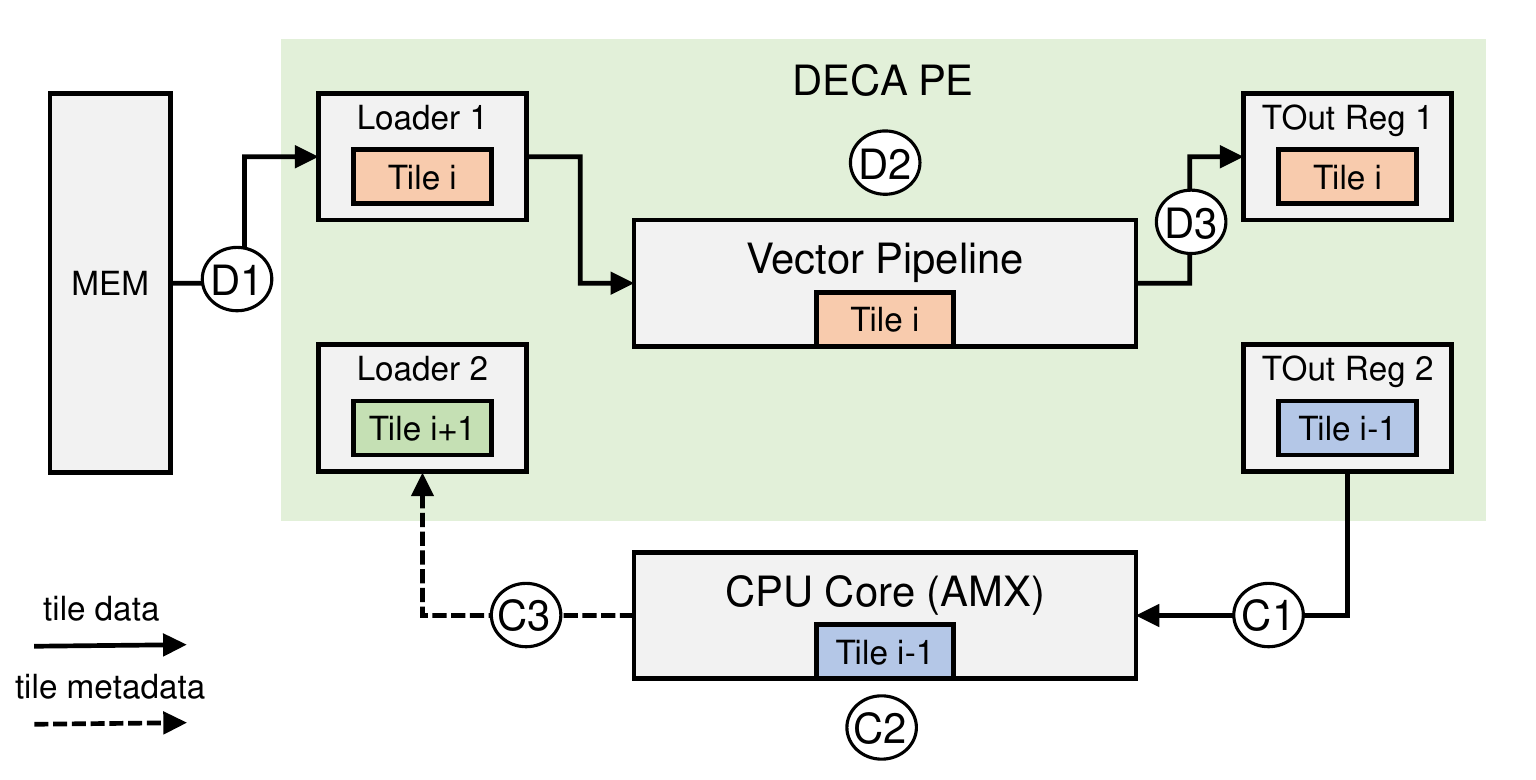}
    \vspace*{-0.5cm}
    \caption{DECA-CPU core cooperative tile processing.}
    \label{fig:accel_cpu_interaction}
    \vspace{-0.3cm}
\end{figure}

To execute GeMMs with high performance, 
we introduce a mechanism that
overlaps 
vector operations in a
DECA with 
AMX 
operations in a CPU core using hardware double buffering.
The design is shown in Figure~\ref{fig:accel_cpu_interaction}. A DECA has two Loader modules
and two TOut registers. A Loader reads a compressed tile from the memory system, which includes three data structures:
the data, a bitmask, and scaling factors. A Loader can also issue prefetches to load a tile in advance.
The journey of a tile involves 
DECA loading it into a Loader (D1 in  Figure~~\ref{fig:accel_cpu_interaction}),
decompressing it in the DECA vector pipeline (D2), and storing it in a TOut register (D3).
Then, the core reads it (C1), uses it to perform the AMX operation (C2), and prompts a Loader (C3) to 
initiate the fetching of the next tile by passing the starting address and the length of the three
data structures of the tile. As shown in the figure,
the double buffers enable overlapping of the operations on two tiles.
While the core is reading and processing Tile \emph{i-1},  DECA reads, processes, and writes
out Tile \emph{i}. After the core finishes  \emph{i-1}, it triggers the fetching of Tile
  \emph{i+1}. 

We explore two options for a CPU core to  communicate with a DECA.
The first one  uses regular stores to the memory-mapped DECA interface;
the second uses  ISA extensions that we describe in Section~\ref{tepl}. 
Using the first approach,  Figure~\ref{lst:mmio} shows the   pseudocode of the core as it
processes  tiles as shown in Figure~\ref{fig:accel_cpu_interaction}. 
The key instructions are those in Lines 4-6. The core uses {\em TLoad} 
(an AMX instruction) to load 
tile $T_{i-1}$
from a DECA TOut register into a tile register TReg$_1$ (Line 4). It then uses this tile in a
{\em TComp} instruction (an AMX instruction that performs a GeMM), saving the output in a tile register
TReg$_2$ (Line 5). Finally, it writes the metadata for tile $T_{i+1}$ (shown as $M_{i+1}$) to 
a memory-mapped register in DECA's Loader2 using a plain store. The write prompts  Loader2 to
initiate the fetch of tile $T_{i+1}$. In parallel with   Lines 4-6, DECA is decompressing $T_{i}$.

\begin{figure}[htb]
\centering
\begin{minipage}[t]{0.45\linewidth}

\begin{mdframed}
    \begin{lstlisting}[escapechar=|]
............
DECA|$_{ldr1}$|  |\(\leftarrow\)| ST M|$_i$|
Fence
TReg|$_{1}$| |\(\leftarrow\)| TLoad T|$_{i-1}$|
TReg|$_{2}$| |\(\leftarrow\)| TComp TReg|$_{1}$|
DECA|$_{ldr2}$|  |\(\leftarrow\)| ST M|$_{i+1}$|
Fence
TReg|$_{1}$| |\(\leftarrow\)| TLoad T|$_i$|
TReg|$_{2}$| |\(\leftarrow\)| TComp TReg|$_{1}$|
............
    \end{lstlisting}
\end{mdframed}
    \caption[]{CPU core pseudocode for store-based DECA invocation.}
    \label{lst:mmio}
    \vspace{-5mm}
\end{minipage}\hfill
\begin{minipage}[t]{0.53\linewidth}

    \begin{mdframed}
        \begin{lstlisting}
............
TReg|$_{1}$| |\(\leftarrow\)| TEPL M|$_{i-1}$|
TReg|$_{2}$| |\(\leftarrow\)| TComp TReg|$_{1}$|
TReg|$_{1}$| |\(\leftarrow\)| TEPL M|$_{i}$|
TReg|$_{2}$| |\(\leftarrow\)| TComp TReg|$_{1}$|
TReg|$_{1}$| |\(\leftarrow\)| TEPL M|$_{i+1}$|
TReg|$_{2}$| |\(\leftarrow\)| TComp TReg|$_{1}$|
............

        \end{lstlisting}
    \end{mdframed}
    \vspace*{-0.4cm}
    \caption{CPU pseudocode for TEPL-based DECA invocation. The architectural tile registers TReg$_1$ and TReg$_2$ get   renamed to  different physical tile registers in each iteration.}
    \label{lst:dal}
    \vspace{-4mm}
\end{minipage}\hfill

\end{figure}

Figure~\ref{lst:mmio} also shows a piece of the previous iteration (Line 2) and of the
subsequent iteration (Lines 8-9). To prevent incorrect memory operation reordering, we 
add a memory fence per iteration. Specifically, the load of tile $T_{i}$ (Line 8) should not execute
before the metadata for $T_{i}$ is written to the control register in DECA's Loader1 (Line 2), which resets TOut Register 1 and initiates the tile fetch from memory.
Since these two instructions do not depend on each other, we place a fence in Line 3. There is
a fence in each iteration.

Unfortunately, this approach is likely to deliver limited performance for two reasons. First, each 
iteration has a fence that prevents cross-iteration overlap.
Second, within an iteration, no instruction overlaps: the instructions in
Lines 4 and 5 have a true dependence, and the store in Line 6 can only perform
the update when it is at the head of the reorder buffer (ROB). 
The execution of all instructions is serialized, as if the core was in-order.
As a result, in every iteration,
the latency of the 
communication between core and DECA (both the load and the store) is fully exposed.

\vspace{-0.15cm}
\subsection{ISA Support for Out-of-Order Invocation}
\label{tepl}

To 
reinstate out-of-order execution 
and hide the core--DECA communication, we propose 
a different approach that relies on an 
extension to the CPU AMX ISA. We call the extension 
{\em Tile External Preprocess and Load} (TEPL). The main idea is to eliminate
the per-iteration fence in Figure~\ref{lst:mmio} by combining, in hardware,
the instructions in Lines 2 and 8 into a single instruction. This instruction
updates the control register of a loader with metadata, triggering a tile fetch, and only returns 
to the core
when DECA has decompressed the tile and stored it in a core tile register (e.g., TReg$_1$).

A TEPL instruction takes as 
arguments a source register with the metadata for a tile,  and a destination core tile register. The metadata is transferred to the DECA to initiate decompression. 
Moreover, the maximum number of TEPL instructions that can execute at any point in time is equal to the total number of DECA Loaders (i.e. two). A structural hazard prevents more TEPLs from executing. This is done to avoid overwriting accelerator invocations, since each DECA loader is able to handle only one tile at a time.

With this design, the code in Figure~\ref{lst:mmio} is rewritten  
as Figure~\ref{lst:dal}. Fences are removed and an iteration has only two instructions
(e.g., Lines 4,5). 
There are no register dependencies between the iterations because
TReg$_1$ and TReg$_2$ are renamed. However, a structural hazard causes the
TEPL in Line 6 to stall until one of the previous two completes. 

A context switch can 
only occur in between two instructions. Hence, the DECA state that needs to be saved 
and restored when a new process attempts to use the DECA is 
only the DECA control registers and LUTs, and not any tile data.

To support these instructions, the core has a \emph{TEPL Queue}  akin to a load-store queue, and two \emph{TEPL execution ports}, each leading to a DECA loader.
As a TEPL instruction $i$ enters the ROB, it is 
deposited in this Queue. 
When its source register is available and  there is a free TEPL execution port, $i$ is issued to the 
DECA.

To attain high performance, TEPLs are issued to the DECA as soon as possible---they 
do not wait until they reach the ROB head. Hence, like a load instruction, they execute speculatively and out-of-order. Invoking a DECA speculatively is always safe, as a DECA does not update memory state. If the core needs to flush the pipeline (e.g., on a branch
misprediction or exception) while
a TEPL instruction is outstanding, the core sends a squash signal to the DECA. At that point, the DECA
aborts any tile operation in progress, no matter the state it is in. The core may safely reissue the 
same TEPL.

Overall, this design hides the communication between the core and DECA. The core 
executes without fences and overlaps the operation on multiple tiles.
TEPLs are not only useful for DECA. A core can potentially use them to communicate with
other DECA-like near-core tile preprocessing accelerators.

\vspace{-0.1cm}
\section{DECA Microarchitecture Design}
\label{sec:microarch}
We now describe the  microarchitecture that enables DECA to sustain high decompression performance
and, at the same time, support a rich set of compression schemes.
For simplicity, in the 
rest of the paper, we assume that DECA's output tile is in BF16 format. DECA can be trivially configured to produce I8 output tiles.

\subsection{DECA Microarchitecture}

\begin{figure}[htb]
    \vspace{-0.15cm}
    \centering
    \includegraphics[width=0.97\linewidth]{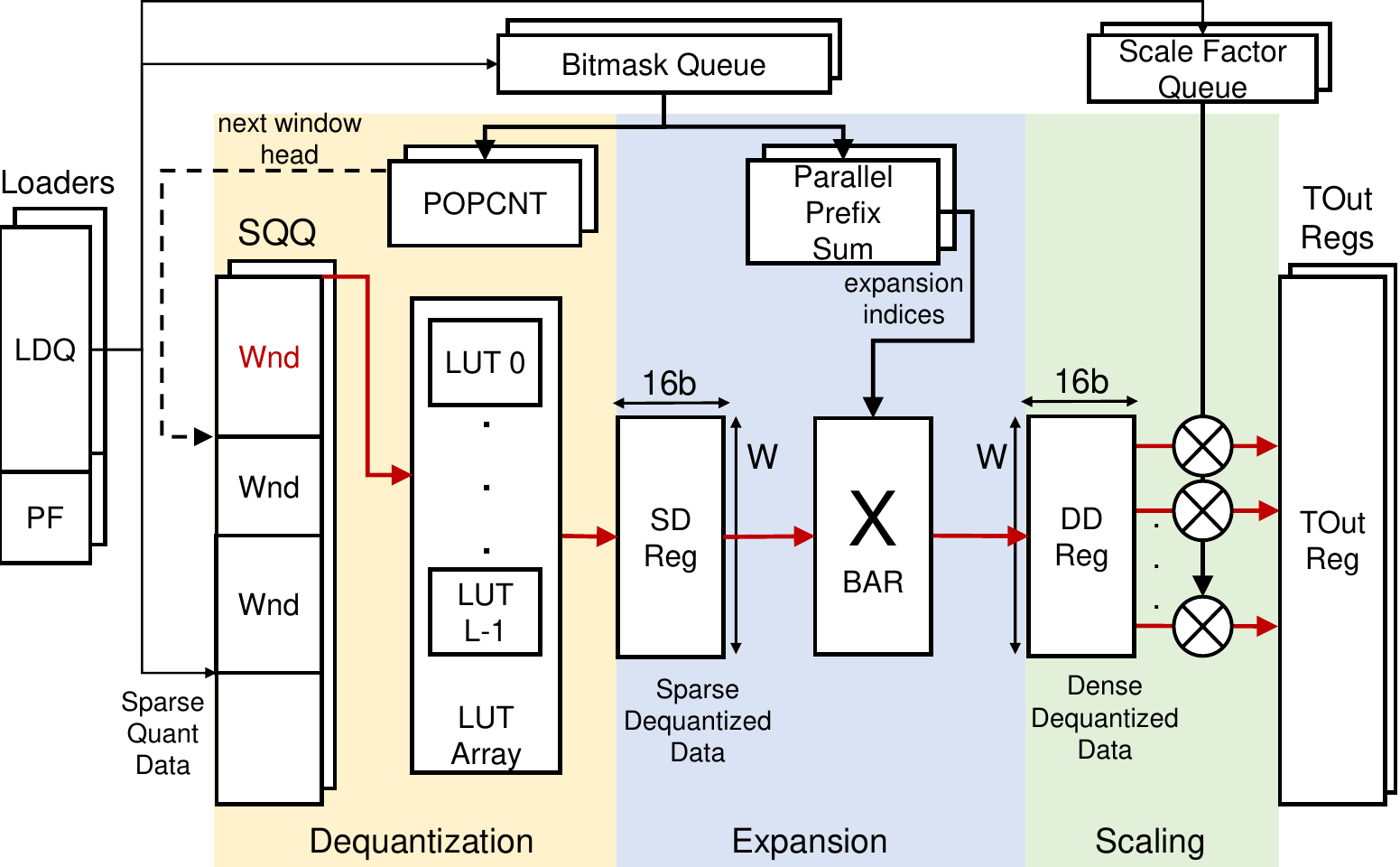}
    \vspace*{-0.3cm}
    \caption{DECA PE  microarchitecture.}
    \label{fig:microarch_overview}
    \vspace{-0.3cm}
\end{figure}

 Figure~\ref{fig:microarch_overview} displays the DECA PE microarchitecture. To understand it,
we describe its multiple components.

\noindent {\bf Accessing Memory.}
DECA has two Loaders, each composed of a \emph{Load Queue} (LDQ) and a
prefetcher (PF). The LDQ
accesses memory to read compressed weights, bitmasks, and scaling factors. The memory address bases and lengths of these structures
are part of the metadata provided by the CPU on DECA invocation. 
When a requested cache line arrives from memory, depending on 
which of the three types of data it contains, 
it is placed in the 
\emph{Sparse Quantized Queue} (SQQ),  \emph{Bitmask Queue}, or  \emph{Scale Factor Queue}. 
The PF observes the address bases and lengths used for a Tile, 
and predicts the ones for future Tiles.
The PF then generates prefetch requests that will bring this data 
to the L2 cache. 
The PF aggressiveness is dynamically adjusted so that a high L2 MSHR occupancy is preserved.

\noindent {\bf  Pipeline Stages.} The pipeline is split into three stages, responsible for dequantization, expansion (i.e., de-sparsification) and scaling. Each stage has its own output register to enable pipelining (SD, DD and TOut). The Dequantization stage reads values from the SQQ, dequantizes them using an array of $L$ Lookup Tables (\emph{LUT Array}), and writes dequantized BF16 values to the {\em Sparse Dequantized} (SD) register. These values are  potentially 
sparse---stored contiguously with zero values skipped. 
The Expansion  stage de-sparsifies data by inserting  zeros in the  positions  indicated by the bitmask. This operation is performed using a crossbar (\emph{XBAR}) that is controlled using expansion indices.
The latter are  generated from the bitmask using the \emph{Parallel Prefix Sum} circuitry. The result is written to the {\em Dense Dequantized} (DD) register, which  contains dense (i.e., with explicit zeros) dequantized data. Finally, if  group  quantization is used,  the Scaling stage applies appropriate scaling to the BF16 values 
by multiplying them with the scaling factors. It then
 writes the final values to the \emph{TOut} register. The critical path is shown  with red arrows in the figure. 

\noindent {\bf  Duplicated Modules.} A DECA PE contains two Loaders and two TOut registers to enable the overlapping of DECA  and CPU  operation. Hence, as shown in Figure~\ref{fig:microarch_overview}, the PE replicates LDQ, PF, the input queues (SQQ, Bitmask queue, and Scale Factor queue), and TOut. One Loader can be supplying data while the pipeline
is processing data that was provided by the other Loader. The bitmask processing circuitry 
mainly performs additions of 1-bit data, and 
is also duplicated so we can hide its latency.
The rest of the pipeline is not duplicated and used by one Loader-TOut pair at a time.

\noindent {\bf  Vector Operations (vOps).} It takes multiple cycles to generate a
decompressed BF16 tile, which always contains 512 BF16 elements. This is because the pipeline
generates output chunks of {\em W} elements at a time, each using one 
DECA Vector Operation (vOp). In the absence of pipeline bubbles, a new chunk is generated every cycle.
A vOp  reads data from the SQQ, executes in the pipeline stages and finally writes W elements to a TOut. vOps exploit pipelining: if a vOp enters the Expansion stage, the next vOp can enter the Dequantization stage. The vOps of a tile are processed in-order and can enter the pipeline as long as (1) their input has arrived from memory and (2) the first pipeline stage is free. 

Without sparsity, a vOp  reads W elements from the SQQ. With sparsity, less than W elements are
needed, since the SQQ does not contain zero values. We refer to the elements that a given vOp needs to read from the SQQ as the vOp's window ({\em Wnd}). 
To determine the 
size of a Wnd, the POPCNT circuitry counts the number of ``1s'' in the bitmask, and determines the end of the current Wnd and the start of the next Wnd. The latter is the next SQQ position from which data will be read into the pipeline.

\noindent {\bf  LUT Array Organization.}
 The DECA dequantization stage supports up to 8-bit quantized numbers, which can represent a maximum of 256 different values. For this reason, each of the $L$ LUTs in the LUT array stores 256 ($2^8$) BF16 values. 
 Dequantizing an 8-bit value corresponds to a lookup using the 8-bit value as the LUT address. DECA contains $L$ LUTs to allow for parallel dequantization of multiple values. Each LUT is internally divided into 4 smaller sub-LUTs, each one with a read port and 64 ($2^6$) entries. If the quantized data bitwidth is 6 bits or less, the 4 sub-LUTs can be used in isolation to enable 4 reads from one 256-entry "big" LUT. For less than 6-bit quantization, some of the LUT entries are redundant and will not be used at runtime.

\noindent {\bf  Bubbles and the Roof-Surface.}
We set the number of   "big" LUTs to $L < W$
to limit DECA's area.
If the Wnd
of a vOp  is larger than $L$ elements, the vOp occupies the Dequantization stage for more than one cycle. This injects one or more \emph{bubbles} in the  pipeline,   which reduce  the vOp throughput. For example, the Wnd of a dense 8-bit quantization scheme is W and, therefore, a
vOp will always require $W/L$ cycles for dequantization. 
Although setting $L$ < $W$ limits the DECA throughput for dense quantization schemes, this is not
a major concern because dense  schemes like BF8\_100\% and MXFP4 require less vector throughput 
(i.e., VOS) in order to escape the vector (VEC) region. This can seen in the BORDs of
Figure~\ref{fig:roofsurface}.

On the other hand, sparser schemes require a higher VOS to escape the VEC-bound region.
Luckily, this is naturally achieved by the DECA pipeline: 
the probability that the Wnd of a vOp is larger than $L$ decreases with sparsity.
Thus, fewer bubbles are introduced for sparse schemes, naturally achieving higher throughput than
their dense counterparts for the same $L$. The same behavior is achieved for lower
bitwidth schemes because  they can perform more than $L$ reads in parallel from the LUT array.

\noindent {\bf Generality and Performance.}
DECA supports quantization formats of 8 bits and lower, group quantization, and unstructured sparsity, which cover most current and likely future model compression schemes.
DECA's design is flexible, since by changing the values
 in its LUT array and/or using different scale factors, it 
  enables the support for a rich set of
formats without redesigning the hardware.
Additionally, individual stages can be skipped if they are unneeded (e.g., quantization without sparsity).
In terms of performance, the main benefit of DECA is that it replaces multiple vector (AVX) instructions by a single vOp that performs the whole decompression: dequantization, expansion, and scaling.
The decreased vOp count   \emph{increases the $AI_{XV}$} (Section~\ref{sec:roofsurface}), moving the points away from the VEC region.
Finally, note that DECA  efficiently dequantizes only the non-zeros, which is hard to do on a CPU with a traditional vector ISA due to data dependent branches during expansion.

\vspace{-0.1cm}

\subsection{Quantitative Microarchitecture Design}
\label{subsec:quant_roofsurface}

In  previous sections, we discussed 
how the Roof-Surface model influenced DECA's design {\em qualitatively}. For example,
it suggested designing a higher-performance accelerator 
by optimizing the $AI_{XV}$, and not just by blindly scaling  the CPU's width and AVX resources. 
We now discuss how it can be used \emph{quantitatively} to dimension DECA's $W$ and $L$ 
parameters and derive a well-balanced design. 

Consider Equation~\ref{equation:roofsurface}.
We should express how the parameters in the equation depend on $W$ and $L$.
In reality, only 
$AI_{XV}$ depends on $W$ and $L$.
\emph{VOS}  is   $c*1*f$, since each of the $c$ CPU cores has one DECA PE that can complete at most one vOp per cycle 
 and operates at the core frequency. On the other hand, the $AI_{XV}$ of  different  kernels depends on  DECA's $W$ and $L$ parameters.
To calculate it, we  need to add-up the number of vOps that are needed per tile and the number of bubbles that are generated per tile.

\begin{table*}[htb]

\centering
\caption{Comparison of DECA with other in/near-core accelerators.}
\vspace{-0.4cm}

\label{tab:related}
\resizebox{0.685\textwidth}{!}{
\begin{tabular}{c|c|c|c|c|c|c|c|c}
\hline
\multicolumn{1}{c|}{Accelerator} & \begin{tabular}[c]{@{}c@{}}Supports\\ Different\\Quantizations\end{tabular} & \begin{tabular}[c]{@{}c@{}}Supports\\ Structured \\ Sparsity\end{tabular} & \begin{tabular}[c]{@{}c@{}}Supports\\ Unstructured\\ Sparsity\end{tabular} & \begin{tabular}[c]{@{}c@{}}GeMM\\ Operation \\ Type\end{tabular} & \begin{tabular}[c]{@{}c@{}}High\\ GeMM\\ Throughput\end{tabular} & \begin{tabular}[c]{@{}c@{}}(I)n or\\ (N)ear\\ Core\end{tabular} & \begin{tabular}[c]{@{}c@{}}Fine-grained\\ Interleaving\\ with the Core\end{tabular} & \multicolumn{1}{c}{\begin{tabular}[c]{@{}c@{}}Changes \\ Required\\ to the Core\end{tabular}} \\ \hline \hline
\textbf{TMUL~\cite{intel_optimization_manual}}          & Limited                                                              & \xmark                                                                        & \xmark                                                                         & Matrix                                                           & \cmark                                                       & I                                                    & \cmark                                                                             & N/A                                                                                            \\ \hline
\textbf{TensorCore~\cite{blackwell}}    & Limited                                                              & 2:4                                                                       & \xmark                                                                         & Matrix                                                           & \cmark                                                       & I                                                      & \cmark                                                                             & N/A                                                                                            \\ \hline
\textbf{RASA~\cite{jeong2021rasa}}          & \xmark                                                                   & \xmark                                                                        & \xmark                                                                         & Matrix                                                           & \cmark                                                       & I                                                    & \cmark                                                                             & Many                                                                                           \\ \hline
\textbf{VEGETA~\cite{jeong2023vegeta}}        & \xmark                                                                   & 2:4,1:4                                                                   & \xmark                                                                         & Matrix                                                           & \cmark                                                       & I                                                    & \cmark                                                                             & Many                                                                                           \\ \hline
\textbf{SAVE~\cite{save}}          & \xmark                                                                   & \cmark                                                                       & \cmark                                                                        & Vector                                                           & \xmark                                                        & I                                                    & \cmark                                                                             & Many                                                                                           \\ \hline
\textbf{SPADE~\cite{gerogiannis23spade}}         & \xmark                                                                   & \cmark                                                                       & \cmark                                                                        & Vector                                                           & \xmark                                                        & N                                                  & \xmark                                                                              & None                                                                                           \\ \hline
\textbf{DECA}     & \cmark                                                                  & \cmark                                                                       & \cmark                                                                        & Matrix                                                           & \cmark                                                       & N                                                  & \cmark                                                                             & Few, reusable                                                                                  \\ \hline
\end{tabular}}
\vspace{-0.35cm}
\end{table*}

The number of vOps per tile 
is  $\#vOps= 512/W$, since each tile has 512 elements and we produce $W$ with a single vOp.
We express the number of bubbles per tile 
as $\#bbl=\#vOps*bpv$, where $bpv$ is
the number of bubbles per vOp. 
Since bubbles can only be generated due to insufficient resources in the Dequantization stage, we use $L_q$ to denote the maximum number of elements that can be dequantized in a cycle.
$L_q$ is equal to $L$ for 8-bit quantization schemes, $2*L$ for 7-bit, and $4*L$ for 6-bit and below. 
Without sparsity, $bpv=ceil(W/L_q) - 1$. With sparsity, the bubble generation is not deterministic, as it depends on the number of nonzeros in a compressed tile.
For a matrix of density  $d$, if we assume that nonzeros are uniformly distributed, then the number of nonzeros in $W$ consecutive matrix elements is a binomial distribution with parameters $W$, $d$.
We  compute the expected number of bubbles 
as:  
\vspace{-0.15cm}
\begin{align*}
bpv &=  \sum\nolimits_{k=0}^{\frac{W}{L_q}-1} k \cdot [F((k+1)L_q; W, d) - F(kL_q; W, d)]
\end{align*}
where $F(i;W,d)$ is the binomial cumulative distribution function. Finally, the $AI_{XV}$ is given by $1/[\#vOps*(1+bpv)]$.

Now we have all  we need  to perform an analytical Design Space Exploration (DSE) using the Roof-Surface model. For example, we can plot the BORDs of different ($W$, $L$) pairs and pick the one that pushes all kernels out of the VEC-bound area at the minimum DECA hardware cost (see Section~\ref{subsec:eval_dse}).

\vspace{-0.1cm}

\vspace{-0.1cm}
\section{Alternatives to DECA for Handling the Decompression Bottleneck}

\label{sec:alternatives}
In Sections~\ref{sec:integration} and~\ref{sec:microarch}, we discussed how   DECA can sustain high decompression performance while maintaining support for a rich set of compression schemes. 
We now discuss the shortcomings of two alternatives to using DECA: scaling the CPU core's vector resources or using other in/near-core accelerator designs.

\noindent
{\bf  1. Traditional scaling of the CPU vector resources.}
Our Roof-Surface analysis of Section~\ref{sec:roofsurface} reveals that, to hide most of the decompression overheads, one would need more than a 4x increase in   vector throughput (VOS). 
Supporting such increase by conventional scaling of a core's vector resources is
very challenging.
One approach would be to increase the number of SIMD AVX vector units by more than 4x. However, as discussed in Section~\ref{sec:roofsurface},   cores are already using 40-80\% of their commit slots.
Hence, such a substantial increase in the number of vector units would   require a major increase in the superscalar core width. This is undesirable, as
a core's area scales quadratically with the superscalar width~\cite{palacharla1997complexity}.

Another approach would be to increase the SIMD AVX vector width. This requires new AVX instructions that operate on multi-cache line operands of at least
2048. However, supporting AVX2048 would require significant ISA and pipeline 
changes  
(e.g., redesigning wider versions of all the vector instructions, new register files, etc.).
In addition, feeding the core with so large vectors would require, at a minimum,
increasing the number of ports in the L1 cache. This would in turn hurt the L1 access latency and the core's cycle time, affecting the core's performance for general-purpose workloads. 
In Section~\ref{sec:evaluation}, we quantitatively compare DECA to these alternatives.

\noindent
{\bf 2. In-core accelerators using matrix operations.}
Traditional matrix units such as the TMUL, and RASA~\cite{jeong2021rasa} cannot deal with compressed tiles.
To avoid the need for tile decompression, 
some in-core accelerator designs~\cite{jeong2023vegeta,peltekis2024demm,blackwell} such as VEGETA~\cite{jeong2023vegeta} augment matrix units with support for specific structured sparsity patterns. Such an approach increases hardware
complexity in the core (e.g., larger matrix unit, more architectural registers, changes in register renaming). 
Further, although this
approach can increase the matrix throughput (\emph{MOS}) by skipping some computations involving zero values, our \emph{Roof-Surface} analysis of Section~\ref{sec:roofsurface} reveals that such an increase is unnecessary for our kernels.
: most of them become bound by memory after escaping the vector-bound region.

Other designs augment the matrix units with native support for more efficient lower bit quantization formats~\cite{jang2024figna,blackwell}. However, such designs require extra hardware to be included in the matrix unit for each one of the supported formats. Further, the hardware needs to be redesigned if a new, previously unseen, quantization format emerges. 
Instead, DECA can support a very rich set of quantization formats without requiring extra hardware for each one of them (i.e., by changing the values in its LUT array and/or  using different scale factors). DECA's flexibility enables support for future quantization formats without redesigning the hardware. 

In principle, all the DECA hardware (i.e., LUT array, expansion and scaling circuitry, etc.) could be integrated in the matrix multiplication unit. However, the decoupled approach of DECA has some important advantages. First, it adds flexibility: the output of the decompressor can also be fed into another accelerator, stored back to memory, or be used for other use-cases.
Second, by attaching the accelerator with its own Loaders at the L2, DECA can more effectively fetch and prefetch data. 
Finally, the CPU core ISA and pipeline changes required are minimal, decreasing the risk of impacting the core's performance in general-purpose workloads.

\noindent {\bf 3. In/near-core accelerators using vector operations.}

\noindent SPADE~\cite{gerogiannis23spade} and SAVE~\cite{save} are accelerators for sparse applications designed to be integrated with CPUs. However, instead of relying on matrix units, they use vector units to execute the actual GeMM. While this approach might work for highly sparse matrices, utilizing the high throughput of matrix units is necessary for the moderately sparse matrices found in machine learning models~\cite{yang2024trapezoid}.

Table~\ref{tab:related} summarizes the unique combination of characteristics enabled by DECA, when compared to other state-of-the-art in/near-core accelerators. 
First, DECA is the first design that offers support for a rich set of quantization schemes combined with structured or unstructured sparsity. At the same time it enables high GeMM throughput, by cooperating with the TMUL matrix units.
Second, through speculative invocation, it is the first near-core accelerator design that enables fine-grained interleaving with the core. Finally, it introduces only few changes to the core's pipeline, which can be reused for other near-core accelerators (Section~\ref{tepl}).

\vspace{-0.2cm}
\section{Methodology}
\label{sec:methodology}

\noindent
{\bf Simulation and System Parameters.}
To evaluate our work, we simulate a 56-core server with SPR-like parameters using an internal simulator based on Sniper~\cite{sniper} with full support for AMX. 
We evaluate the DDR5-based and the HBM-based designs with about 260GB/s and 850GB/s achievable memory bandwidth, respectively.
We extend the simulator with: (1) DECA PEs, 
and (2) a TEPL queue and ports to support TEPLs in the core pipeline.
Both cores and DECA PEs run at 2.5GHz.
Our baseline PE is dimensioned with W=32 and L=8, but we also evaluate other options in Section~\ref{subsec:eval_dse}.

\noindent
{\bf Software and DECA Control Code Generation.}
We use the Intel Libxsmm compressed GeMM kernels  (Section~\ref{subsec:libxsmm}) as our software baseline. 
To invoke DECA, we modify the libxsmm JIT compiler by replacing the AVX decompression sequence with TEPL instructions.

To evaluate the effectiveness of DECA for compressed GeMMs in isolation, we implement a large cascade of Fully Connected (FC) layers (without other types of layers) and use Parlooper~\cite{georganas2023harnessing} for loop parallelization. The weight matrices in those layers have $\approx$250 million parameters, similar to the large FC layers of Llama-2-70B. Libxsmm and Parlooper are already integrated in the Intel Tensor Processing Primitives (TPP) Framework~\cite{georganas2021tensor}, which supports end-to-end Llama-2  and OPT  inference on CPUs. Hence, we use TPP as is for software-only LLM inference, and by invoking the TEPL-augmented libxsmm kernels for inference with DECA. We test batch sizes 
of 1-16. Our simulator is compatible with all frameworks. 

\noindent
{\bf  Compression Schemes.} In our evaluation, we refer to BF16, BF8, and MXFP4 as
Q16, Q8, and Q4.
We limit the compression schemes we evaluate to these, since these are the ones for which
libxsmm already includes 
support. 
We also evaluate unstructured sparsity with weight density ranging from 50\% to 5\% for Q16 (only sparsity) and Q8 (quantization plus sparsity). The Q4 sparse kernels are currently not included in libxsmm, so we don't have reference data to compare with DECA performance. For end-to-end Llama-2-70B and OPT-66B  inference, 
the uncompressed Q16 baseline, Q16 with 50\% density (Q16\_50\%) and Q8\_100\% do not fit in the 64GB of HBM. Hence, we simulate a larger HBM capacity for those schemes.
Note that the Q4 performance is also representative of INT4 compression schemes with scaling factors  such as AWQ~\cite{awq}.

\noindent
{\bf   Area estimation.} We estimate the area of our proposed 
DECA design with W=32 and L=8. 
For the memory structures (e.g., LDQ and SQQ), registers, and   LUT array,
we use  CACTI~\cite{cacti}.  
For the crossbar and for the  BF16 multipliers, we use numbers from~\cite{cakir2015modeling}
and ~\cite{zhang2019new}, respectively.
We then use~\cite{stillmaker2017scaling} to scale down the numbers to 7nm.
We estimate the total area for 56 DECA PEs to be around 2.51 $mm^2$. The 
Loaders, SQQs, Bitmask queues, Scale Factor queues, and 
TOut registers consume about 55\% of DECA's area, the LUT array consumes 22\%, while the 
rest consumes 23\%.
Given that the total die area of a 56-core SPR is around 1600 $mm^2$~\cite{diesize}, the DECA area overhead is less than 0.2\%.

\vspace{-0.2cm}
\section{Evaluation}
\label{sec:evaluation}

\subsection{DECA for Compressed GeMMs} \label{subsec:eval_gemms}

\begin{figure}[htb]
    \centering

      \includegraphics[width=0.94\linewidth]{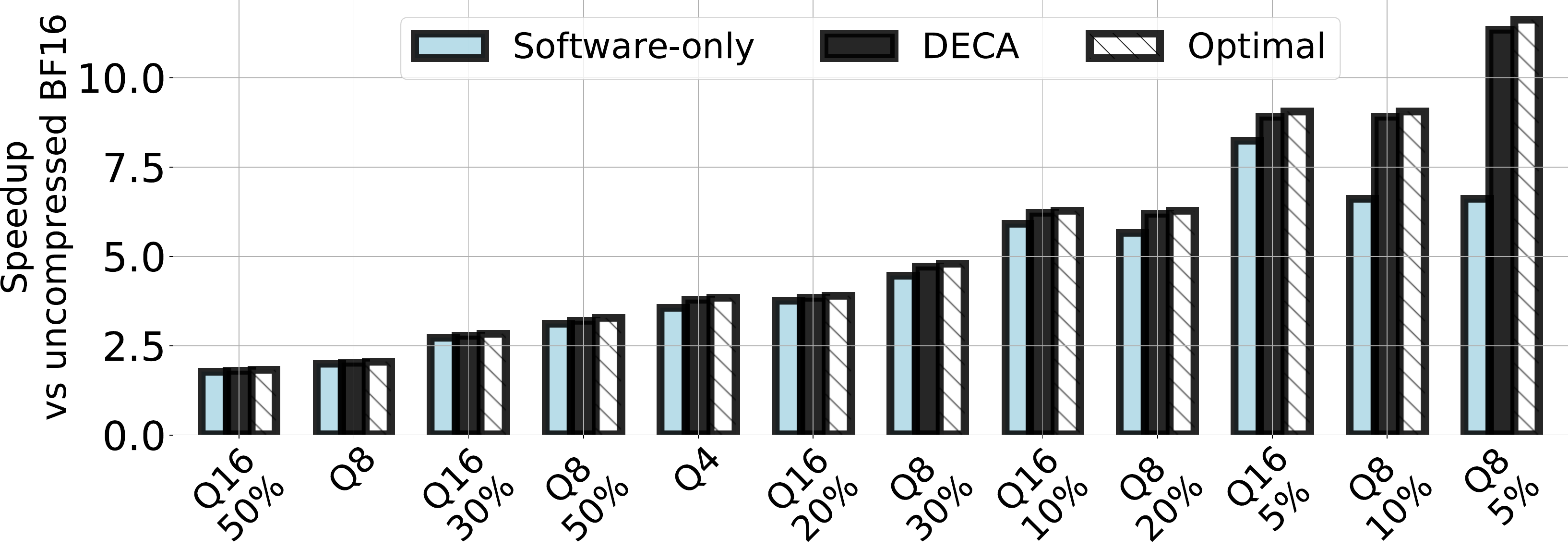}
      \vspace*{-0.3cm}
  \caption{Compressed GeMM speedup for DDR and N=1.}
  \label{fig:main_ddr}
  \vspace{-0.3cm}
\end{figure}

\begin{figure}[htb]
    \centering

      \includegraphics[width=0.94\linewidth]{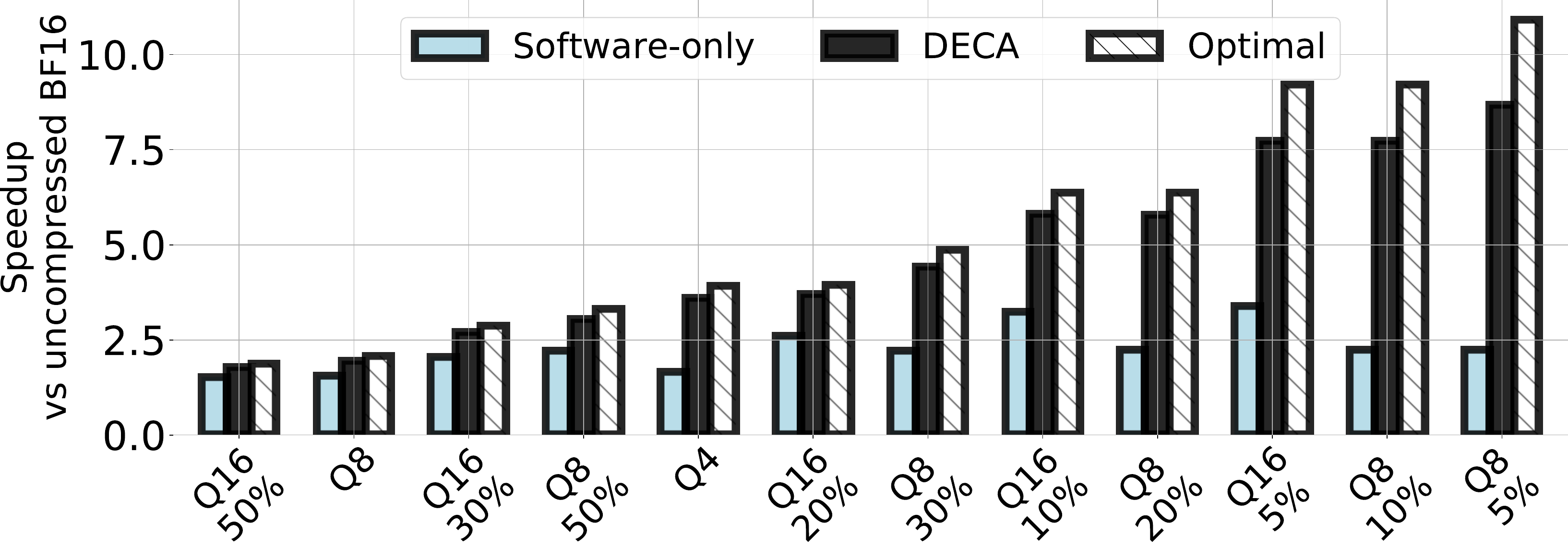}
      \vspace*{-0.3cm}
  \caption{Compressed GeMM speedup for HBM and N=1.}
  \vspace{-0.15cm}
  \label{fig:main_hbm}
\end{figure}

Figures~\ref{fig:main_ddr} and~\ref{fig:main_hbm} show, for different compression schemes, the speedups of the libxsmm software solution ({\em Software-only}) and of DECA over the baseline uncompressed BF16 scheme. We also add the {\em Optimal} speedup from the \emph{roofline} model, which assumes that all VEC overheads are hidden. The
compression schemes appear with increasing compression factor.
We show results for N=1.

For the DDR setting (Figure~\ref{fig:main_ddr}), DECA offers speedups over software only for  high compression factors. This is expected since, according to the BORD in Figure~\ref{fig:bored_ddr}, only high compression factors are VEC-bound. 
The speedups reach 1.7$\times$.
For the HBM setting (Figure~\ref{fig:main_hbm}), DECA offers speedups for almost all the compression schemes. This is because,  as shown by the BORD in Figure~\ref{fig:bored_hbm}, almost all schemes are VEC-bound. The speedups reach 4.0$\times$.
In both DDR and HBM, the performance of DECA is near-optimal,   revealing that the VEC overheads are successfully hidden. We repeated this analysis for batch sizes 
of up to N=16 and observed similar results.

\begin{figure}[h]
    \centering
      \includegraphics[width=0.9\linewidth]{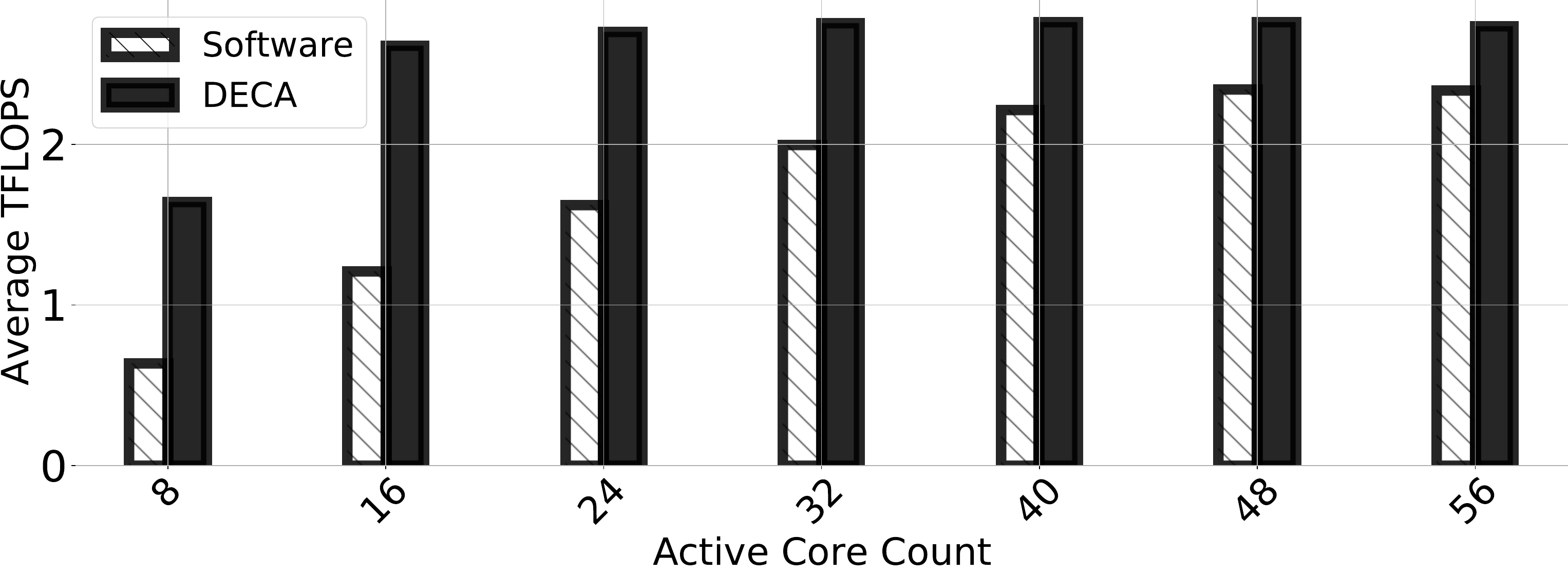}
      \vspace*{-0.35cm}
  \caption{TFLOPS across all compressions for DDR and N=4.}
  \label{fig:ddr_core_sweep}
  \vspace{-0.3cm}
\end{figure}

DECA-augmented cores are much more capable at vector processing than conventional cores.
Figure~\ref{fig:ddr_core_sweep} compares the performance of both types of cores
for the DDR setting with N=4, averaged across all the compression schemes. The 
figure compares different core counts: 8, 16, ...56. We see, e.g., that
16 DECA-augmented cores achieve higher performance than 56 conventional cores.
The extra cores can either be freed-up for other workloads that do not
consume much memory bandwidth, or power-gated to save energy.

To provide further insights into the performance of the system with software only and the one with DECA, Table~\ref{tab:utilizations} displays the 
percent utilization of the memory bandwidth, of the TMUL, and of either the CPU's AVX units  or DECA. 
Since performance is proportional to the utilization of the TMUL, the table shows that
the system with DECA has much higher performance than software-only. 
Further, since the operations of 
the three components overlap, the one with the highest utilization ends up being the bottleneck. In the software-only system, for almost all of the densities, the 
bottleneck is the AVX vector units. This observation validates the Roof-Surface prediction.
With DECA, the memory is much better utilized, leading to direct performance improvements.
 Note that, although sparser kernels take less time to execute, the utilization of DECA remains fairly constant. As explained in Section~\ref{sec:microarch},  
DECA naturally achieves higher throughput for sparse schemes.

\begin{table}[htb]
\vspace{-0.2cm}
\centering
\footnotesize
\caption{Component utilization for Q8, N=1, and HBM.}
\vspace{-0.3cm}
\label{tab:utilizations}
\begin{tabular}{|c||ccc||ccc|}
\hline
                 & \multicolumn{3}{c||}{\textbf{Software-only}}                                                      & \multicolumn{3}{c|}{\textbf{With DECA}}                                                          \\ \hline \hline
\textbf{Density} & \multicolumn{1}{l|}{\textbf{MEM}} & \multicolumn{1}{l|}{\textbf{TMUL}} & \textbf{AVX} & \multicolumn{1}{l|}{\textbf{MEM}} & \multicolumn{1}{l|}{\textbf{TMUL}} & \textbf{DECA} \\ \hline
\textbf{100\%}   & \multicolumn{1}{c|}{74\%}           & \multicolumn{1}{c|}{14\%}            & 50\%            & \multicolumn{1}{c|}{93\%}           & \multicolumn{1}{c|}{18\%}            & 75\%           \\ \hline
\textbf{50\%}    & \multicolumn{1}{c|}{66\%}           & \multicolumn{1}{c|}{20\%}            & 88\%            & \multicolumn{1}{c|}{92\%}           & \multicolumn{1}{c|}{28\%}            & 71\%            \\ \hline
\textbf{20\%}    & \multicolumn{1}{c|}{35\%}           & \multicolumn{1}{c|}{20\%}            & 89\%            & \multicolumn{1}{c|}{91\%}           & \multicolumn{1}{c|}{53\%}            & 63\%            \\ \hline
\textbf{5\%}     & \multicolumn{1}{c|}{19\%}           & \multicolumn{1}{c|}{20\%}            & 89\%           & \multicolumn{1}{c|}{73\%}           & \multicolumn{1}{c|}{79\%}            & 87\%            \\ \hline
\end{tabular}
\vspace{-0.25cm}
\end{table}

\begin{figure}[h]
 \vspace{-0.15cm}
    \centering

      \includegraphics[width=0.94\linewidth]{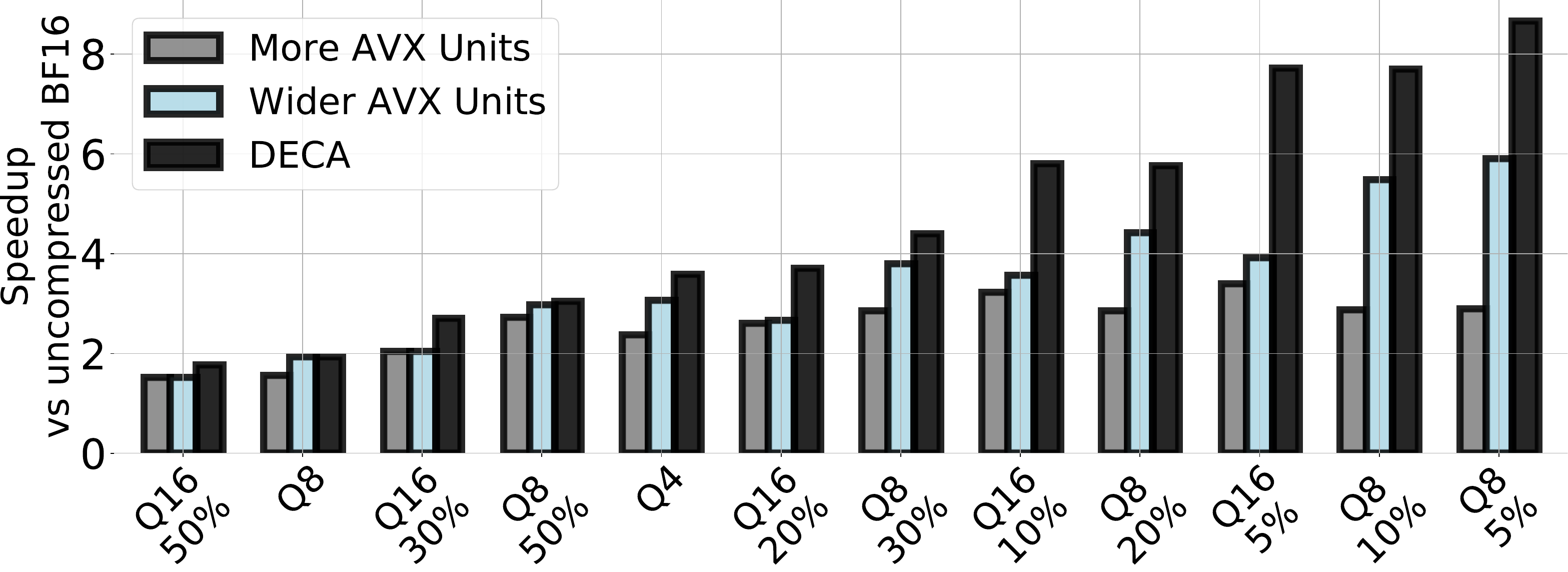}
      \vspace*{-0.35cm}
  \caption{DECA vs traditional vector scaling for HBM \& N=1.}
  \label{fig:strawmen}
  \vspace{-0.45cm}
\end{figure}

Figure~\ref{fig:strawmen}   compares DECA with the alternative of scaling the CPU core's vector resources as a method to alleviate the decompression overhead. We compare a DECA-augmented core to a core with: (1)  4$\times$ more vector AVX units ({\em More AVX Units}) or (2)  4$\times$ wider AVX units ({\em Wider AVX Units}). We optimistically model the wider AVX2048 units by removing the dynamic instructions from 3 out of 4 iterations of the decompression loop. Since we do not modify the system cache line, each AVX2048 memory operation is executed as 4 cache-line sized operations. 
For the non-DECA systems, we do 
not scale the superscalar width of the core or the number of L1 ports  since, as explained in Section~\ref{sec:alternatives}, such changes are prohibitive. From the figure, we see that  the performance of conventional vector scaling methods is  far below DECA's performance.

\subsection{Design Space Exploration with Roof-Surface} \label{subsec:eval_dse}

\begin{figure}[h]
    \vspace{-0.3cm}
    \centering
  \subfloat[No DECA]{%
      \includegraphics[width=0.24\textwidth]{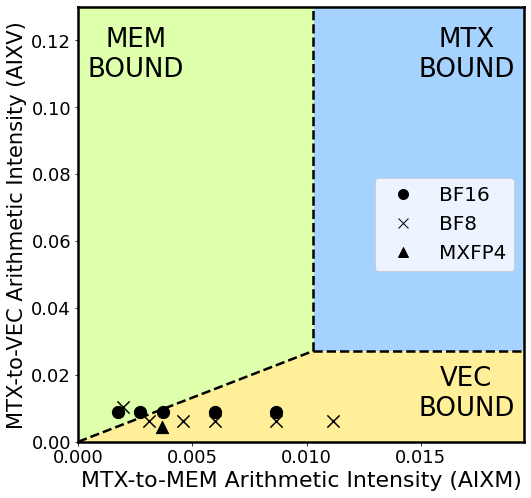}
      \label{fig:dse_cpu}
      }
  \subfloat[Different-sized DECA]{%
      \includegraphics[width=0.24\textwidth]{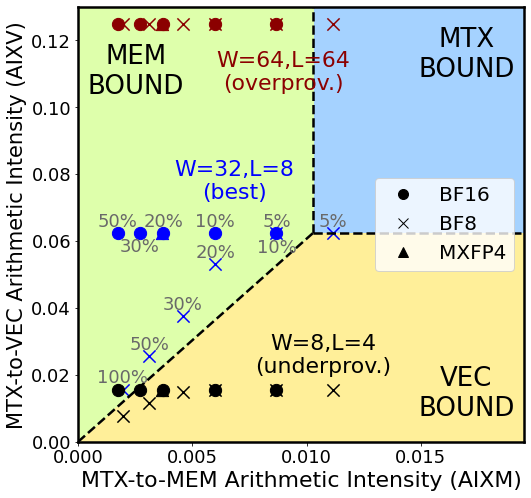}
      \label{fig:dse_best}
      }
    \vspace*{-0.3cm}
  \caption{HBM BORDs with no DECA and with different-sized DECAs.}
  \label{fig:dse_rooflines}
  \vspace{-0.25cm}
\end{figure}

The DECA W and L parameters determine how fast DECA can  decompress, but too large values may increase area without real benefit. To this end, we use the
\emph{Roof-Surface} to examine the performance  for different \{W,L\} pairs. To dimension DECA, we pick the smallest \{W,L\} pair for which the predicted performance saturates (i.e., all the kernels are predicted  \emph{not} to be VEC-bound anymore). According to our model, this  value pair is \{W=32,L=8\}. In Figure~\ref{fig:dse_rooflines}, we compare the BORDs
for the HBM SPR system 
without DECA (a) and with DECA (b) with different \{W,L\} sizes: \{W=8,L=4\} (underprovisioned), \{W=32,L=8\} (best),
and \{W=64,L=64\} (overprovisioned).

We observe that, in comparison with CPU, DECA has a smaller vector operations per second ($VOS$) parameter, since its VEC-bound region is larger. However, DECA decreases the number of vector operations needed per matrix operation (i.e., it increases $AI_{XV}$) as discussed in Section~\ref{sec:microarch}. The underprovisioned DECA with \{W=8,L=4\} is unable to push the kernels out of the VEC-bound region. The overprovisioned one with \{W=64,L=64\} pushes them out, but more than needed. We simulate the performance of these pairs to validate the model's accuracy. We find that the DECA-best system  is 2$\times$ faster than the DECA-underprovisioned
one. The DECA-overprovisioned system is less than 3\% faster than the 
DECA-best one. At the same time, DECA-best is much cheaper than DECA-overprovisioned:
it has 8$\times$ fewer LUTs and half the W.
Overall, the Roof-Surface model accurately captures the dynamics of the matrix-vector-memory interaction and can guide microarchitectural decisions.

\vspace{-0.1cm}

\subsection{Analysis of DECA Integration and TEPLs} \label{subsec:eval_integ}

We now evaluate  different decisions we made regarding
DECA's integration with a core. We start with
a base configuration where DECA reads compressed tiles from the LLC (bypassing L2), writes decompressed tiles in the L2 for the core to read, and is invoked using normal loads, stores, and fences. Then, we progressively enhance it to: (1) allow the accelerator to read compressed weights from the L2 and use the L2 prefetcher ({\em +Reads L2}), (2) use its own prefetcher instead of the L2 prefetcher ({\em +DECA prefetcher}), (3) write to the TOut Regs instead of to the L2 ({\em +TOut Regs}), and (4) use TEPL instructions instead of loads, stores, and fences ({\em +TEPL (DECA)}).

\begin{figure}[htb]
\vspace{-0.2cm}
    \centering
      \includegraphics[width=0.95\linewidth]{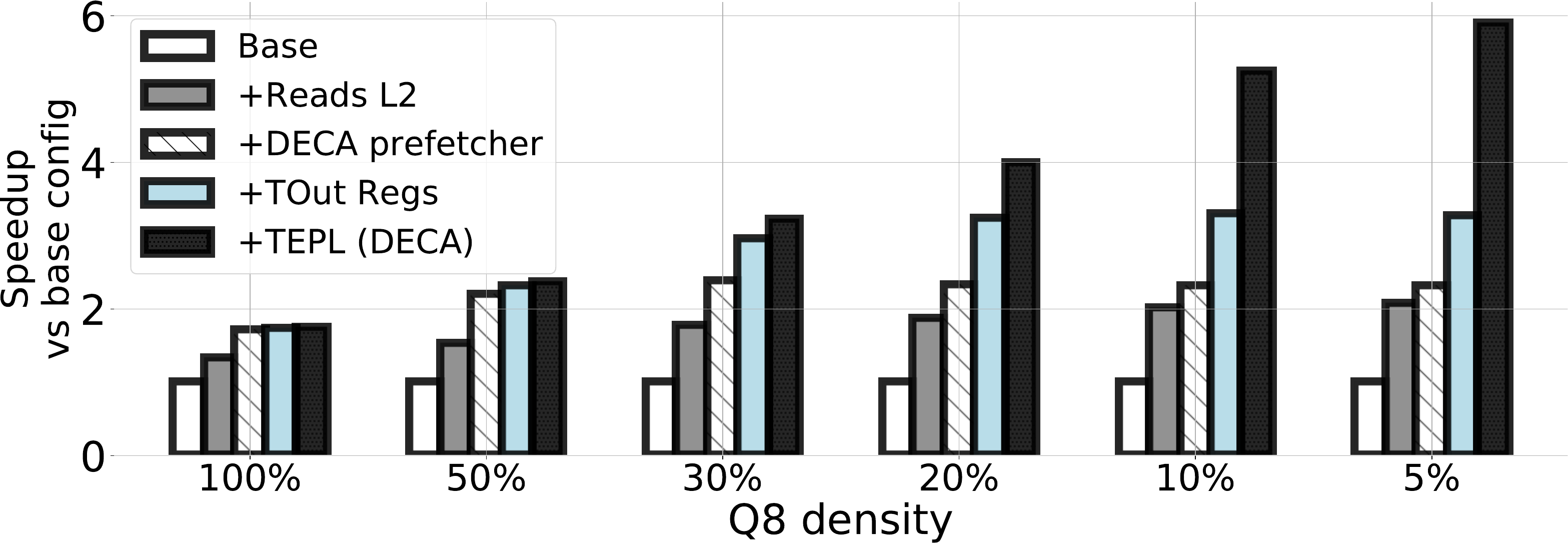}
      \vspace*{-0.4cm}
  \caption{DECA integration features for HBM and N=4.}
  \label{fig:ablation}
   \vspace{-0.36cm}
\end{figure}

Figure~\ref{fig:ablation}
shows the speedups over the base design that the progressive application of these
optimizations obtains for Q8 with different densities.
We see that {\em +Reads L2} improves performance for all densities. The benefit comes from the L2 hardware prefetcher already available in the system, which fetches future tiles, hiding the memory and LLC access latencies. {\em +DECA prefetcher} further improves performance by
using the DECA prefetcher rather than the default L2 one.
{\em +TOut Regs} and {\em +TEPL (DECA)} 
reduce or hide the DECA-core communication latency and are necessary for out-of-order invocation.
Specifically, {\em +TOut Regs} enables the core to directly fetch data from DECA, instead of taking the longer path through the L2. Further,
{\em +TEPL (DECA)}  overlaps  communication with computation, effectively hiding the former. 
We see that the effectiveness of {\em +TOut Regs} and {\em +TEPL (DECA)} 
increases   as the density decreases. This is because
DECA takes less time to process a lower density tile, while the overhead of communication 
with the core  remains constant. Thus, for lower densities, 
the communication cost gets more exposed.
Note that TEPLs are very effective for low-density models: for 5\% density, they double the performance.

\vspace{-0.3cm}
\subsection{DECA for LLM Inference} \label{subsec:eval_llm}

\vspace{-0.1cm}

Lastly, we show the performance benefit of DECA for LLM next token generation (including the non-GeMM stages).
Table~\ref{tab:benchmarks} shows the next token latencies of the Llama2-70B and OPT-66B models, respectively, on SPR with HBM, for 128 input tokens, 128 output tokens, batch sizes 1 and 16, and different compression schemes.
We compare  software decompression  ({\em SW}) with our proposal ({\em DECA}).
As explained, we simulate  the uncompressed baseline BF16 model assuming a larger HBM size. 
We see that DECA reduces the next token time by 1.6$\times$--2.6$\times$ over {\em SW}. This translates into a  2.5$\times$--5.0x speedup over the uncompressed base model.
We observed similar results for shorter/longer token sequences.

\begin{table}[htb]
\vspace{-0.25cm}
\centering
\caption{Llama2-70B/OPT-66B next-token latency (ms)}
\vspace{-0.45cm}
\label{tab:benchmarks}
\resizebox{0.9\linewidth}{!}{
\begin{tabular}{|l|llll|llll|}
\hline
              & \multicolumn{4}{c|}{\textbf{Batch Size = 1}}                                                                                                                                                                                                                                                                                                 & \multicolumn{4}{c|}{\textbf{Batch Size = 16}}                                                                                                                                                                                                                                                                            \\ \hline
              & \multicolumn{1}{l|}{\textbf{\begin{tabular}[c]{@{}l@{}}BF16\\ 100\%\end{tabular}}} & \multicolumn{1}{l|}{\textbf{\begin{tabular}[c]{@{}l@{}}MXFP4\end{tabular}}} & \multicolumn{1}{l|}{\textbf{\begin{tabular}[c]{@{}l@{}}BF8\\ 20\%\end{tabular}}} & \multicolumn{1}{l|}{\textbf{\begin{tabular}[c]{@{}l@{}}BF8\\ 5\%\end{tabular}}} & \multicolumn{1}{l|}{\textbf{\begin{tabular}[c]{@{}l@{}}BF16\\ 100\%\end{tabular}}} & \multicolumn{1}{l|}{\textbf{\begin{tabular}[c]{@{}l@{}}MXFP4\end{tabular}}} & \multicolumn{1}{l|}{\textbf{\begin{tabular}[c]{@{}l@{}}BF8\\ 20\%\end{tabular}}} & \textbf{\begin{tabular}[c]{@{}l@{}}BF8\\ 5\%\end{tabular}} \\ \hline
              & \multicolumn{8}{c|}{\textbf{Llama2-70B}}                                                                                                                                                                                                                                                                                                                                                                                                                                                                                                                                                                                                                              \\ \hline
\textbf{SW}   & \multicolumn{1}{l|}{192.3}                                                        & \multicolumn{1}{l|}{124.6}                                                        & \multicolumn{1}{l|}{98.1}                                                      & \multicolumn{1}{l|}{98.1}                                                     & \multicolumn{1}{l|}{211.2}                                                        & \multicolumn{1}{l|}{139.1}                                                        & \multicolumn{1}{l|}{116.2}                                                      & 115.8                                                     \\ \hline
\textbf{DECA} & \multicolumn{1}{l|}{-}                                                             & \multicolumn{1}{l|}{68.3}                                                         & \multicolumn{1}{l|}{50.5}                                                       & \multicolumn{1}{l|}{40.7}                                                      & \multicolumn{1}{l|}{-}                                                             & \multicolumn{1}{l|}{82.3}                                                         & \multicolumn{1}{l|}{66.5}                                                       & 56.8                                                     \\ \hline
              & \multicolumn{8}{c|}{\textbf{OPT-66B}}                                                                                                                                                                                                                                                                                                                                                                                                                                                                                                                                                                                                                              \\ \hline
\textbf{SW}   & \multicolumn{1}{l|}{178.5}                                                        & \multicolumn{1}{l|}{116.9}                                                        & \multicolumn{1}{l|}{91.2}                                                      & \multicolumn{1}{l|}{91.0}                                                     & \multicolumn{1}{l|}{203.9}                                                        & \multicolumn{1}{l|}{132.3}                                                        & \multicolumn{1}{l|}{111.4}                                                      & 110.8                                                     \\ \hline
\textbf{DECA} & \multicolumn{1}{l|}{-}                                                             & \multicolumn{1}{l|}{60.8}                                                         & \multicolumn{1}{l|}{45.0}                                                       & \multicolumn{1}{l|}{35.6}                                                      & \multicolumn{1}{l|}{-}                                                             & \multicolumn{1}{l|}{81.8}                                                        & \multicolumn{1}{l|}{64.3}                                                      & 55.5                                                      \\ \hline

\end{tabular}}
\vspace{-0.3cm}
\end{table}

\vspace{-0.15cm}
\section{Other Related work}
\vspace{-0.1cm}

\label{sec:related}

\noindent
{\bf Decoupled Accelerators.}
A variety of stand-alone decoupled accelerators
that target sparsity in  ML and scientific applications have been proposed~\cite{zhang2016cambricon,lu2019efficient,gerogiannis2024hottiles,scnn,chen2022regraph,hedge19extensor,srivastava20tensaurus,gondimalla2019sparten,han2016eie,adiletta2023characterizing,piuma}. Other decoupled accelerators 
rely on quantization
~\cite{zhu2024mega,ryu2022bitblade,jang2024figna}. Recently, accelerators for attention are also becoming popular~\cite{wang2021spatten,kachris2024survey,lu2021sanger,ham2021elsa,ham20203}.
Decoupled accelerators come with large area-power budgets~\cite{jeong2023vegeta}, and suffer from data movement overheads~\cite{gerogiannis23spade}. For those reasons, CPU-integrated accelerators have been proposed~\cite{gerogiannis23spade,jeong2023vegeta,save,graphite,jeong2021rasa,nassif2022sapphire}. DECA falls in this line of works.
We discussed the shortcomings of other in/near-core accelerators in Section~\ref{sec:alternatives}.

\noindent
{\bf  Cooperative Vector-Matrix Processing.}
A variety of architectures include heterogeneous matrix and vector units, whose interaction could be modeled  with the \emph{Roof-Surface} model. Examples include the Tandem processor~\cite{tandem},
the AWS Trainium~\cite{trainium,trainium2}, 
the TPU~\cite{tpuv23}, and GPUs with their Tensor Cores and SIMT Cores.

\noindent
{\bf  Utility of a DECA-inspired decompression engine for GPUs.}
Similar to the TMUL, the GPU Tensor Cores support only limited quantization formats and do not support unstructured sparsity. For this reason, GPU kernels such as Flash-LLM~\cite{xia2023flashllm} adopt a similar approach to  libxsmm: compressed data is decompressed through software and fed to the Tensor Cores. Although effective, Flash-LLM puts pressure on the L1/shared memory of the SMs, preventing full TensorCore/HBM utilization. We thus believe that DECA-inspired decompression engines could also be useful for GPUs. NVIDIA recently introduced the TMA accelerator~\cite{luo2024benchmarking} for supplying data from memory to Tensor Cores. Augmenting TMA with DECA-inspired decompression capabilities is an interesting future direction.

\vspace{-0.2cm}

\section{Conclusion}
\vspace{-0.1cm}

To improve LLM inference in advanced CPU platforms
with in-core GeMM engines and HBM, this paper made  three  contributions:
the   {\em Roof-Surface} performance model, the {\em DECA} 
near-core ML-model decompression accelerator, and the TEPL ISA extension for out-of-order accelerator invocation.
Our evaluation shows that DECA effectively accelerates compressed GeMMs and LLM inference.


\bibliographystyle{ACM-Reference-Format}
\bibliography{references}

\end{document}